\documentclass[10pt,twocolumn,twoside] {IEEEtran}
\usepackage{graphicx}
\usepackage{booktabs}
\usepackage{amsmath}
\usepackage{amssymb}
\usepackage{amsthm}
\usepackage{midfloat}
\usepackage{bm} 
\usepackage[dvipsnames]{xcolor}
\usepackage{cite}
\usepackage{cite}
\usepackage{algorithm}
\usepackage{caption}
\usepackage{cuted}
\usepackage{enumitem}
\usepackage[caption=false,font=footnotesize]{subfig}

\newcommand{\real}[1]{\mathrm{Re}\{#1\}}
\newcommand{\imag}[1]{\mathrm{Im}\{#1\}}

\newcommand{\E}{\mathbb E}
\newcommand{\nextline}{\\ \nonumber &\times}

\begin{document}

\title{ADMM-Net for Communication Interference Removal in Stepped-Frequency Radar}
\author{Jeremy~Johnston, Yinchuan~Li, Marco Lops, \emph{Fellow}, \emph{IEEE}, Xiaodong Wang, \emph{Fellow}, \emph{IEEE}
\thanks{
J. Johnston is with the Electrical Engineering Department, Columbia University, New York, NY 10027, USA (e-mail: j.johnston@columbia.edu).
Y. Li is with the School of Information and Electronics, Beijing Institute of Technology, Beijing 100081, China, and the Electrical Engineering Department, Columbia University, New York, NY 10027, USA (e-mail: yinchuan.li.cn@gmail.com).
M.~Lops is with the Department of Electrical Engineering and Information Technologies, Universit\`a di Napoli ``Federico II'', Via Claudio, 21 - I-80125 Naples (Italy) (e-mail: lops@unina.it).
X. Wang is with the Electrical Engineering Department, Columbia University, New York, NY 10027, USA (e-mail: wangx@ee.columbia.edu).
}
}
\maketitle

\begin{abstract} Complex ADMM-Net, a complex-valued neural network architecture inspired by the alternating direction method of multipliers (ADMM), is designed for interference removal in super-resolution stepped frequency radar angle-range-doppler imaging. Tailored to an uncooperative scenario wherein a MIMO radar shares spectrum with communications, the ADMM-Net recovers the radar image---which is assumed to be sparse---and simultaneously removes the communication interference, which is modeled as sparse in the frequency domain owing to spectrum underutilization. The scenario motivates an $\ell_1$-minimization problem whose ADMM iteration, in turn, undergirds the neural network design, yielding a set of generalized ADMM iterations that have learnable hyperparameters and operations. To train the network we use random data generated according to the radar and communication signal models. In numerical experiments ADMM-Net exhibits markedly lower error and computational cost than ADMM and CVX.
\end{abstract}


\section{Introduction}

The use of radar in civilian life has expanded---e.g. automotive radar, remote sensing, and healthcare applications---meanwhile next-generation communications systems have begun to encroach upon spectrum once designated solely for radar use \cite{6967722}. In response, the U.S. Department of Defense declared an initiative \cite{7485100} to spur research on algorithm and system designs that allow radars to cope with the changing spectral landscape. Subsequently, several system design motifs have materialized in the area of radar/communication coexistence \cite{8828016}.

\emph{Coordinated coexistence} methods enable coexistence through system co-design and information sharing. Joint-design of the radar waveform and communication system codebook may be cast as an optimization problem to, e.g., maximize the communication rate subject to constraints on the radar SNR \cite{zheng2017joint}. In a radar-centric co-design, the radar waveform might be forced to lie in the null-space of the channel between the radar and communication devices, based on channel state information provided either externally, or by the radar's own means of channel estimation \cite{sodagari2012projection}. In some proposals the coexisting systems communicate with a data fusion center, which uses the shared information to configure each system in a way that optimizes the performance of the ensemble \cite{7953658}. \emph{Uncoordinated coexistence} methods, on the other hand, seek to minimize interference absent external information; for example, spectrum occupancy measurements can inform real-time adjustments to the transmit waveform, e.g. center frequency \cite{8574912}, and beamforming can mitigate directional interference \cite{deng2013interference}. 

In uncoordinated interference removal, thresholding or filtering can be effective if the interference is much stronger than the desired signal, although runs the risk of inadvertently distorting the desired signal. Parametric methods estimate the parameters of a statistical signal model, via either subspace methods or optimization. Greedy methods, e.g. CLEAN and matching pursuit, project the recording onto an interference dictionary and iteratively build up an interference estimate by finding the dictionary component with the highest correlation, removing that component from the recording, and repeating the process until a stopping criterion is met. If the received interference is concentrated in narrow regions along some dimension, e.g. time, frequency, or physical space, and hence is sparse in a known dictionary, convex relaxation methods such as $\ell_1$-minimization can be effective \cite{li2019multi,8777296}. In multiple measurement processing the interference matrix may be a low-rank, paving the way for nuclear norm minimization \cite{R19}. In this vein, the present paper addresses an uncoordinated scenario where the interference is \emph{sparse} in a known domain. In particular, we show that the stepped-frequency radar waveform's ``frequency-hopping" property imposes on the interference a certain structure that can be leveraged.

\begin{figure*}[t]
\centering
	\subfloat[][]{\includegraphics[width=76mm]{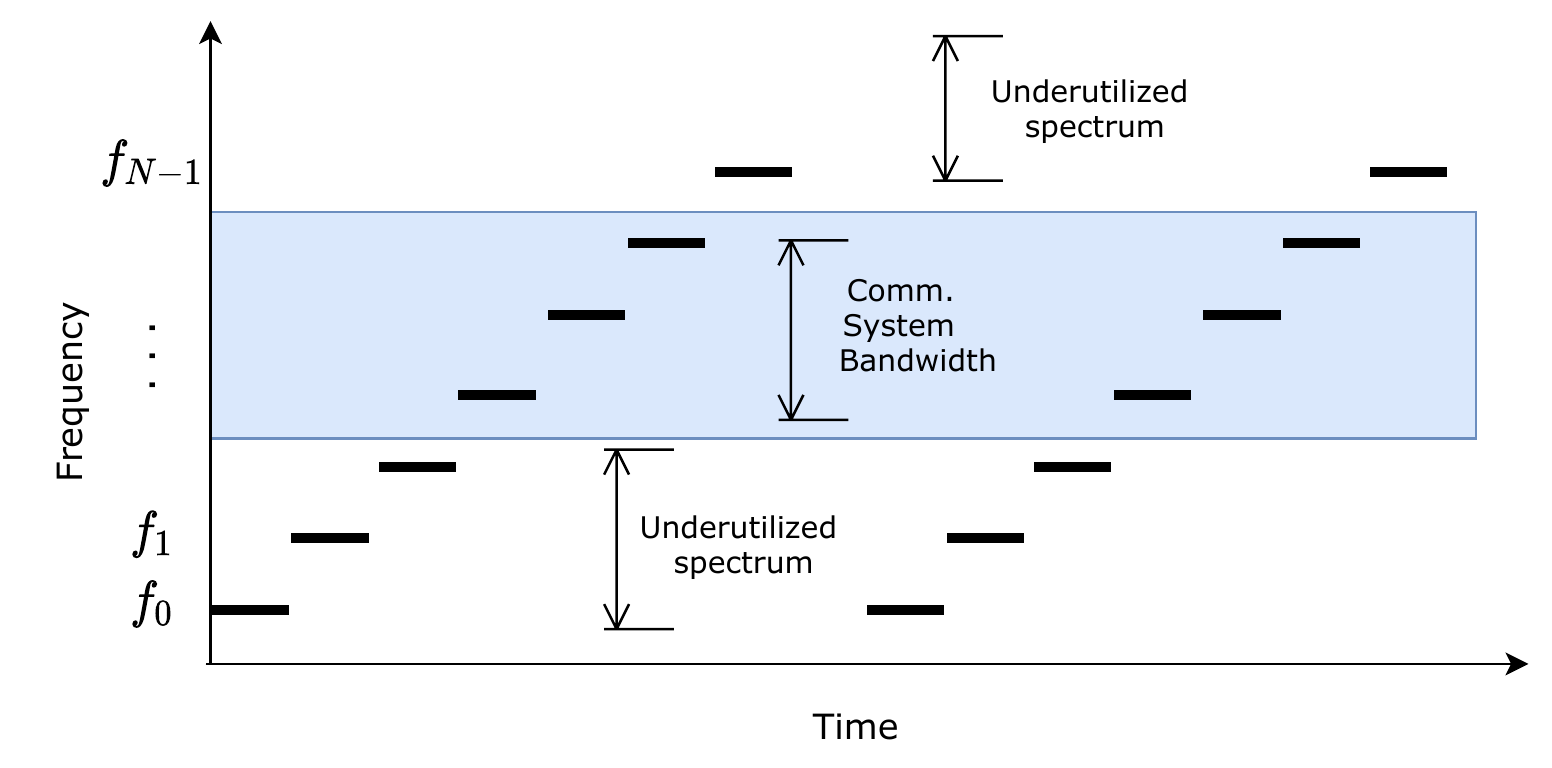}}
	\subfloat[][]{\includegraphics[width=90mm]{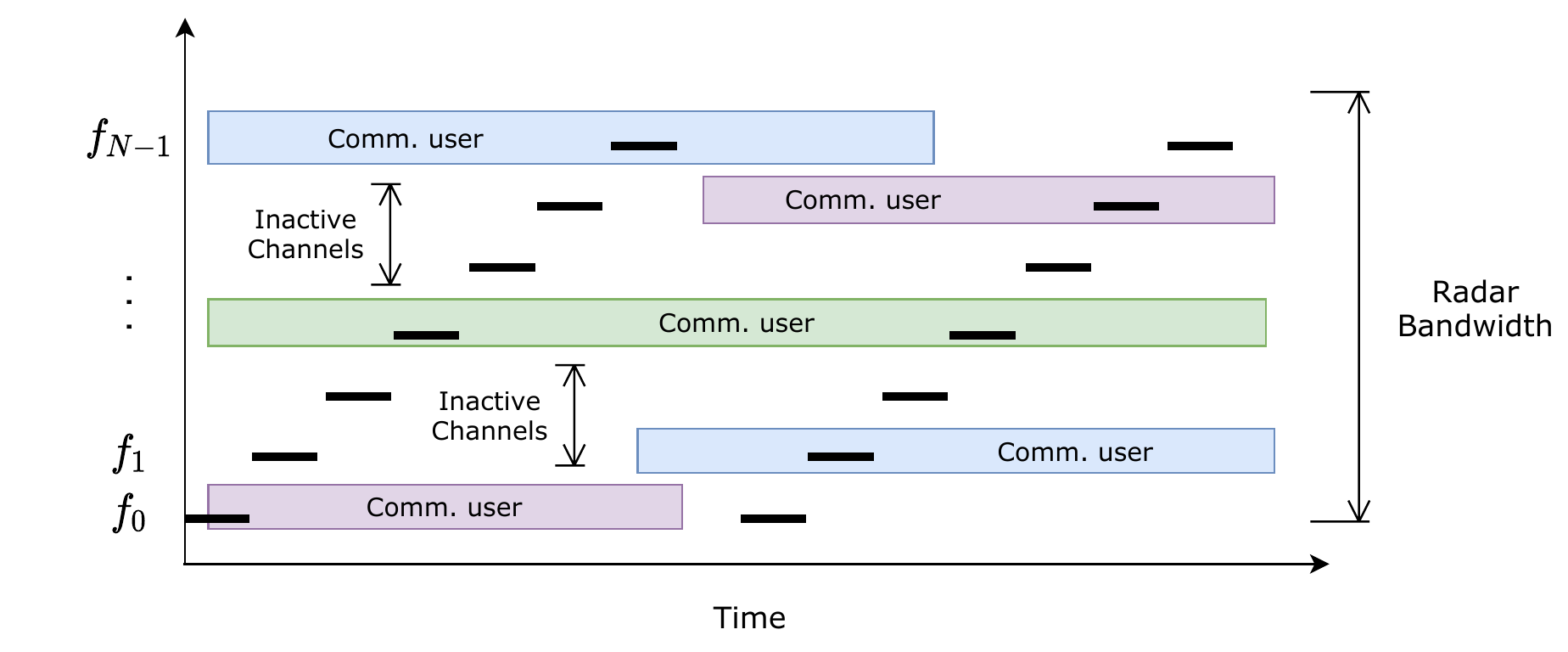}}
	\caption{Frequency occupation versus time for two representative spectrum sharing scenarios. The black strips indicate the spectrum occupied by the radar system; the colored strips indicate the spectrum occupied by the communication system. Only the overlapping regions cause interference to the radar.}
	\label{fig:spectrum}
\end{figure*}

Neural networks are attractive for interference suppression, as they can learn an inverse mapping to recover a signal from corrupted measurements \cite{rs01654,7952561}. So-called ``black box" neural networks may generalize well, but provide only empirical, rather than theoretical performance guarantees, and moreover they neglect the corpus of model-based signal recovery theory and algorithms which exploit prior knowledge to devise computational procedures tailored to the problem \cite{5456165}. Iterative algorithms, grounded in either optimization or statistics, are among the most computationally efficient for signal recovery, but their performance hinges on the careful selection of hyperparameters, whose favorable values are generally problem-dependent. From one point of view, deep unfolding, the approach we adopt in the sequel, automates hyperparameter selection by casting cross-validation as instance of deep learning.

In the deep unfolding \cite{hershey2014deepunfold} framework, a given iterative algorithm inspires a neural network design. Typically the network's forward pass is computationally equivalent to a handful of algorithm iterations, a fraction of that required for the original algorithm to converge, yet the trained network may outperform the original algorithm. In the design, the algorithm's update rules are cast as a block of network layers whose forward pass emulates one full iteration of the algorithm, and whose learnable parameters correspond to a chosen parameterization of the update rules---which may include, for example, algorithm hyperparameters as well as entries of a matrix involved in an update rule. A number of such blocks---possibly augmented, e.g. by a convolutional layer \cite{yang2016admmnet}, in order to increase learning capacity---are sequenced to form the network.  Network training, typically via gradient-based optimization, employs data either gathered from the field or randomly generated according to a priori models, and hence adapts the algorithm to the problem at hand. The layer parameters can be initialized either as prescribed by the algorithm, or even randomly---in one study, an unfolded vector-approximate message passing (VAMP) network randomly initialized learned a denoiser identical to the statistically matched denoiser \cite{7934066}. Algorithms previously considered for deep unfolding include the iterative shrinkage thresholding algorithm (ISTA) \cite{li2020multi}, robust principal components analysis (RPCA) \cite{8836615}, and ADMM \cite{yang2016admmnet}. Applications span those of iterative optimization itself, e.g. wireless communication \cite{8642915}, medical imaging \cite{8836615}, and radar \cite{8768138}. 

In this paper, we design an ADMM-Net which simultaneously recovers a super-resolution angle-range-doppler image \cite{4770164} and removes communication interference. We target an uncooperative spectrum sharing scenario wherein the radar is considered the primary function and the communications utilize portions of the shared spectrum. In the proposed multi-frame radar processing architecture, the stepped-frequency radar transmits a series of simple pulse trains to obtain a set of low-resolution radar measurements, with which the ADMM-Net is able to synthesize an image. Although the total radar bandwidth is large ($\sim1$ GHz), by virtue of the pulse-by-pulse processing only the communication signals that spectrally overlap with a given pulse interfere with the pulse's return. Moreover, communication signals tend to be sparse in the frequency domain (Fig. \ref{fig:spectrum}), owing to periods of low activity or otherwise underutilized spectrum \cite{1391031,5783948}. Consequently, the interference manifests as sparse noise in the radar measurements. 
This motivates an optimization problem which jointly recovers the image and removes the interference. The problem's corresponding ADMM equations, in turn, undergird the design of a neural network, the training of which is tantamount to optimizing a handful of ADMM iterations over their associated hyperparameters and matrices. Important for radar processing, the network processes complex-valued measurements, and does so in a manner consistent with ADMM. Training data sets are randomly generated via the signal model. Experiments indicate the trained ADMM-Net recovers more accurate images than ADMM and CVX at a fraction of the computational cost.

The remainder of the paper proceeds as follows. First, we develop a model of the radar and communication signals and formulate an optimization problem to jointly recover the radar image and interference (Section \ref{sec:signalmodelprobform}). We then derive the problem's ADMM recursion (Section \ref{sec:admmalgo}) and design an ADMM-Net by unfolding the complex-valued ADMM equations into a real-valued neural network (Section \ref{sec:complexadmmnet}). Finally, numerical simulations (Section \ref{sec:simulations}) compare the performance of ADMM-net to that of ADMM and CVX.

\section{Signal Model \& Problem Formulation}
\label{sec:signalmodelprobform}
A stepped-frequency MIMO radar illuminates a sparse scene in the presence of interfering communication signals which are sparse in the frequency domain. The radar undertakes pulse-by-pulse processing over multiple measurement frames, and the joint image recovery-interference removal task is cast as an optimization problem. 

\subsection{Signal Model}

\subsubsection{MIMO Radar Signal}
\label{sec:mimoradarsignal}
Consider a frequency-stepped pulsed MIMO radar with $N_{T}$ transmitters and $N_{R}$ receivers. Each of the transmitted waveforms $u^p, {p=1,\dots,N_{T}}$ has duration $T$ seconds and the waveforms are assumed to be approximately mutually incoherent (see (\ref{eq:pulsexcorr})). The scene is illuminated by $N_d$ trains of $N$ pulses; within the $m$th train, the $n$th pulse emitted by the $p$th transmit antenna is given by
\begin{align}
\label{eq:st}
s^p(m,n,t)= u^p(t - nT_r - mNT_r)\exp{(j2\pi f_n t)},
\end{align}
where $t$ is continuous time, $0 \leq m \leq N_d-1$, $0\leq n \leq N-1$, $1 \leq p \leq N_{T}$; $f_n =  n \Delta f + f_0$ where $f_0$ is the lowest carrier frequency, and $N\Delta f$ is the overall bandwidth. Each pulse echo recording length is $T_r \gg T$ seconds, which is thus the pulse repetition interval (PRI). A complete observation consists of $N_dN$ PRIs.

We consider a scene of $L$ scatterers with scattering coefficients $x_i$ and radial velocities $v_i$. 
The signal received by the $q$th receiver, $q = 1,\dots,N_R$, is
\begin{align}
\label{eq:r_n(t)}
r^{q}(n,t) &= \sum_{m=0}^{N_d -1} \sum_{p=1}^{N_{T}} \sum_{i=1}^L x_i s^p(m,n,t - \tau_i^{pq}(t)),
\end{align}
where
\begin{align}
\label{eq:taui}
\tau_i^{pq}(t) &= \frac {2 v_i}ct+ \tau_i + \delta_i^{p} + \varepsilon_i^{q}
\end{align}
is the $i$th scatterer's delay; $\delta_i^{p}$ and $\varepsilon_i^{q}$ are the marginal delays due to array geometry associated with antenna pair $(p,q)$; and $\tau_i$ is the absolute round-trip delay observed by a reference antenna pair during the first PRI. We assume the velocities are constant throughout the series of sweeps.
\begin{table}[!htb]
\caption{Index of MIMO radar variables}
\label{tab:radarparams}
\centering
\begin{tabular}{cccccc}  
\toprule
Symbol&Definition\\
\midrule
$N$ & No. frequency steps\\  
$N_d$ & No. sweeps \\
$N_T$ & No. transmitters\\
$N_R$ & No. receivers \\
$f_0$ & Start frequency \\
$\Delta f$ & Frequency step size\\
$f_n$ & $f_0 + n \Delta f, \ 0\leq n \leq N-1$ \\
$u^p$ & Transmitter $p$'s pulse envelope \\
$s^p$ & Transmitter $p$'s waveform \\
$r^q$ & Radar return at receiver $q$ \\
$T$ & Pulse duration (all transmitters) \\
$T_r$ & Pulse-repetition interval \\
$t$ & Continuous fast-time, absolute \\
$m$ & Sweep index \\
$n$ & Pulse index within sweep \\
$i$ & Scatterer index \\
$L$ & No. scatterers \\
$x_i$ & scattering coeff.\\
$\tau_i^{pq}$ & absolute delay, $(p,q)$ Tx/Rx pair \\
$\delta_i^{p}$ & marginal delay, $p$th Tx \\
$\varepsilon_i^{q}$ & marginal delay, $q$th Rx  \\
$\tau_i$ & absolute delay, reference Tx/Rx pair \\
$\overline\tau_i(k)$ & delay offset, $k$th range cell \\
$v_i$ & radial velocity \\
$\boldsymbol\theta_i$ & direction coordinates \\
\bottomrule
\end{tabular}
\end{table}
We further make the following assumptions:
\begin{itemize}
\item The range variation throughout the series of sweeps is negligible with respect to the range resolution of each pulse: \[\frac {2v_i N_dNT_r}c \ll T.\]
\item The array element spacing is much smaller than the range resolution granted by the overall transmitted bandwidth:
\begin{align}
\label{eq:eltspacing}
\delta_i^{p} + \varepsilon_i^{q} \ll \frac 1{N\Delta f}.
\end{align}
\end{itemize}
Since the pulse is unsophisticated, $T\Delta f \simeq 1$; hence (\ref{eq:eltspacing}) implies $\delta_i^{p} + \varepsilon_i^{q} \ll T$, whereby
\begin{align}
u^p(t-\tau_i^{pq}(t)) \simeq u^p(t -\tau_i).
\end{align}
In (\ref{eq:r_n(t)}) the term $\exp{(-j2\pi n\Delta f (\delta_i^{p} + \varepsilon_i^{q}))}$ can be neglected since, by (\ref{eq:eltspacing}), $n\Delta f(\delta_i^p + \varepsilon_i^q) \ll 1$, $n=0,1,\dots,N-1$.
With these assumptions, (\ref{eq:r_n(t)}) becomes
\begin{align}
\label{eq:r_n(t)doppler}
r^{q}(n,t) &= \sum_{m=0}^{N_d -1} \sum_{p=1}^{N_{T}} \sum_{i=1}^L x_i \exp{(-j2\pi f_0 (\delta_i^p + \varepsilon_i^q))} \\ \nonumber &\times u^p(t - nT_r - mNT_r - \tau_i)  \exp{(j2\pi f_n (t-\tau_i-\frac {2 v_i}c t))}.
\end{align}

\subsubsection{Communication Signal}
\label{sec:commsignal}

Suppose there are $N_c$ carriers that spectrally overlap with the radar band, with center frequencies $f_i^C$ and bandwidths $B_i$, $i=0,1,\dots,N_c-1$. Here, the term ``carrier" refers to any communication transmission within the radar band; e.g. a particular block of subcarriers within a communication band, the aggregate transmission over a communication band, etc.
The received communication signal has the form
\begin{align}
s_c(t) &= \sum_{i = 0}^{N_c-1}  g_i(t)\exp{(j2\pi f^C_i t)},
\end{align}
where $g_i$ represents the information signal transmitted over carrier $i$ and is a zero-mean random process whose power spectral density $G_i$ satisfies
\begin{align}
G_i(f) = 0 \ \mathrm{if} \ |f| > \frac {B_i} 2.
\end{align}

Applicable scenarios lie between two extremes. At one (Fig. \ref{fig:spectrum}(a)), the total radar bandwidth overlaps with multiple communication carriers and the radar frequency step is on the order of the communication carrier bandwidth. For example, stepped frequency radars may have a step size of 20 MHz \cite{4276877}, while the maximum LTE bandwidth is 20 MHz \cite{3gppLTE} and in sub-6GHz 5G the maximum channel bandwidth is 100 MHz \cite{8752473}. At the other (Fig. \ref{fig:spectrum}(b)), the radar overlaps with a single communication carrier. The carrier comprises sub-channels sized on the order of the radar frequency step-size that are assigned to opportunistic communication users. For example, 5G employs channels with bandwidths in the hundreds of megahertz to a few gigahertz \cite{7894280}, and stepped frequency radars often have a sweep bandwidth on that order. In any case, the key property that enables the radar to coexist is that significant portions of spectrum tend to be underutilized \cite{1391031} \cite{8752473}. In light of this, the interference induced by the active portions can be mitigated.

As a concrete example, to be evaluated in Section \ref{sec:simulations}, consider an uplink SC-FDMA system, such as was specified in the 5G New Radio standard released by 3GPP in December 2017. Suppose the system bandwidth consists of $N_s$ subcarriers with uniform spacing $\Delta f^C$ and every $K\in\mathbb Z^+$ consecutive subcarriers are grouped into channels with center frequencies $f_i^C = f_0^C + iK\Delta f^C, \ 0\leq i \leq N_c-1$, where $f_0^C$ is the start frequency, each channel has bandwidth $K\Delta f^C$, for a total of $N_c = \lfloor N_s/K \rfloor$ channels. Users are assigned one or more channels over which to transmit. The signal transmitted over channel $i$ has the form
\begin{align}
\label{eq:giscfdma}
g_i(t) &= \sqrt{\gamma_i} h_i\sum_{n_c=-\infty}^{\infty} \sum_{k=0}^{K-1} a_{ik}(n_c) u_C(t-n_cT_c)\nextline\exp{\left[j2\pi (f_i^C + k\Delta f^C) t\right]},
\end{align}
where:
\begin{itemize}
\item $\gamma_i$ is the power level assigned to channel $i$.
\item $h_i \sim \mathcal {CN}(0,\beta)$ are i.i.d. channel fading coefficients. A block fading channel model is assumed and $K$ is chosen such that $K\Delta f^C$ equals the coherence bandwidth ($\sim 0.5 \ \mathrm{MHz})$ \cite{8871348}. Therefore each channel $i$ is characterized by a single fading coefficient $h_i$ that is statisticaly independent of all other channels. The variance $\beta$ accounts for additional user-dependent effects (e.g. path loss and log-normal shadowing) \cite{8871348}. For simplicity, we assume $\beta$ is the same for all users.
\item $\{a_{ik}(n_c)\in \mathbb C: 0\leq k \leq K-1, \ 0\leq i \leq N_c-1, \ n_c \in \mathbb Z\}$ are random variables representing the transmitted symbol sequence, comprising the data and cyclic prefix, with $a_{ik}(n_c)$ transmitted on subcarrier $k$ of channel $i$ during the $n_c$th data block. In SC-FDMA the transmitted symbols $a_{ik}(n_c), \ k=0,\dots,K-1$ are the isometric discrete Fourier transform (DFT) coefficients of the original data symbol sequence. We assume the original data symbols adhere to a memoryless modulation format.
\item $T_c$ is the block duration (cyclic prefix plus data); for example, in 5G $\Delta f^C \sim 15$ kHz, so $T_c \sim  1/(15 \mathrm{\ kHz}) = 66 \ \mu s$.
\item
\begin{align}u_C(t) \triangleq \begin{cases}
\sqrt {\frac {1}{T_c}} & 0 \leq t \leq T_c \\
0 & \mathrm{otherwise}
\end{cases}\end{align}
is the normalized pulse envelope.
\end{itemize}

\subsection{Signal Processing at Radar Receiver Side}
\label{sec:rsp}
Receiver $q$'s recording of the $n$th pulse has the form
\begin{align}
\label{eq:f(t)}
\chi^{q}(n,t) &= r^{q}(n,t) + s_c(t) + e(t), \quad 0 < t < N_dNT_r,
\end{align}
where $e(t)$ is additive white gaussian noise (AWGN). Each pulse return is divided into $\lfloor T_r/T \rfloor$ range gates of size $T$ seconds, a range interval of $\frac {cT}2$ meters, centered at times $t_k = kT + \frac T2, \ k = 0,\dots, \lfloor T_r/T \rfloor-1$. The $q$th receiver's recordings are projected onto the $p$th transmit waveform shifted to range cell $k$, i.e. onto the functions $\{s^p(m,n,t - t_k): 0 \leq m \leq N_d-1, 0\leq n \leq N-1, 1\leq p \leq N_{T}, \ 0\leq k \leq \lfloor T_r/T \rfloor-1\}$, to obtain the output sequence $y^{q}(m,n,p,k)$, namely
\begin{align}
\label{eq:projections}
y^{q}(m,n,p,k) &= \left\langle\chi^{q}(n,t), s^p(m,n,t - t_k)\right\rangle\\
\label{eq:yn(k)}
&\triangleq y_R^{q}(m,n,p,k) + y^{q}_C(m,n,p,k) + \overline e(m,n,p,k),
\end{align}
where $\langle y_1(t),y_2(t)\rangle \triangleq \int_{-\infty}^\infty y_1(t)y_2^*(t)dt$ and the terms $y^{q}_R, y^{q}_C,$ and $\overline e$ are the projections of the radar echoes, the communication signal, and AWGN, respectively.
This operation is equivalent to matched filtering each of the $N$ echo recordings and sampling the output at times $t_k$ \cite{zheng2018overlaid}. Next, we develop models for the terms in (\ref{eq:yn(k)}). 

\subsubsection{Radar signal component}
We have
\begin{align}
&y_R^{q}(m,n,p,k) = \langle r^q(n,t), s^p(m,n,t - t_k)\rangle \\
\label{eq:dopplerapprox}
&\simeq  \sum_{p'=1}^{N_{T}}\sum_{i=1}^L x_i \exp{(-j2\pi f_0 (\delta_i^{p'} + \varepsilon_i^q))} R_{u^{p'}u^p}(t_k-\tau_i) \\ \nonumber &\times \exp{(-j2\pi f_n(\tau_i + \frac {2 v_i}c (nT_r + mNT_r) - t_k))} ,
\end{align}
where $R_{uv}(\tau) \triangleq \langle u(t),v(t-\tau) \rangle$, and we have used the fact that $\{u^p(t - nT_r - mNT_r - t_k)\}_{m=0}^{N_d-1}$ is orthogonal along $t$. 
The approximation in (\ref{eq:dopplerapprox}) assumes the target velocities are small enough that the target position is constant within a single PRI. Since each $u^p$ has duration $T$, the autocorrelation $R_{u^pu^p}$ has a duration of approximately $2T$; therefore we assume
\begin{align}
\label{eq:Ru_assumption}
R_{u^pu^p}(\tau) \simeq \begin{cases} 1 & |\tau| < T/2 \\ 0 & |\tau| > T/2\end{cases}.
\end{align}
We also assume the waveforms are incoherent, i.e.
\begin{align}
\label{eq:pulsexcorr}
R_{u^{p'}u^p}(\tau) \simeq 
\begin{cases}
R_{u^pu^p}(\tau) & \mathrm{if} \ p=p' \\
0 & \mathrm{if} \ p \neq p'
\end{cases}, \ \tau \in \left[-\frac T 2, \frac T 2\right].
\end{align}
This could be achieved, for example, through time-domain multiplexing, which would require increasing the illumination period in order to achieve a given maximum unambiguous range.
Define $\mathcal{I}_k \triangleq \{i:|\tau_i-t_k| < T/2\}$,
the indices of the scatterers that belong to range cell $k$. Applying (\ref{eq:Ru_assumption}) and (\ref{eq:pulsexcorr}), (\ref{eq:dopplerapprox}) becomes
\begin{align}
\label{eq:ypqRfinal}
y_{R}^q(m,n,p,k) 
& = \sum_{i\in \mathcal{I}_k} x_i \exp{(-j2\pi f_0 (\delta_i^{p} + \varepsilon_i^q))} \\ \nonumber &\times \exp{\left[-j2\pi(n\Delta f (\tau_i -t_k) + f_0 \frac{2v_i}c mNT_r) \right]} \\ 
\nonumber &\times \exp{\left[-j2\pi(f_n \frac{2v_i}c nT_r + n \Delta f \frac{2v_i}c mNT_r) \right]},
\end{align}
where we have absorbed $\exp{(-j2\pi f_0 (\tau_i - t_k))}$ into $x_i$.

In general the Tx/Rx array elements are distributed on a plane and the delays $\delta_i^p = \delta_i^p(\boldsymbol\theta)$ and $\varepsilon_i^q = \varepsilon_i^q(\boldsymbol\theta)$ are functions of the scatterer's angular coordinates $\boldsymbol\theta \in \mathbb R^2$, e.g. azimuth and elevation, relative to the array plane. We consider a generic array response matrix $\mathbf H \in \mathbb C ^ {N_{T}\times N_{R}}$ where 
\begin{align}
[\mathbf H(\boldsymbol\theta)]_{pq} \triangleq \exp{(-j2\pi f_0 (\delta_i^{p}(\boldsymbol\theta) + \varepsilon_i^q(\boldsymbol\theta))}
\end{align}
and let $\mathbf h \triangleq \mathrm{vec}\left(\mathbf H\right) \in \mathbb C ^ {N_{T}N_{R}}$. 

We define steering vectors for the intra- and inter-frame time scales: for intra-frame, the range steering vector $\mathbf r(\tau,v) \in \mathbb C ^ N$ where
\begin{align}
\label{eq:rangesteer}
[\mathbf r(\tau,v)]_n &\triangleq \exp{\left[-j2\pi(n\Delta f \tau + f_n \frac{2v}c nT_r) \right]};
\end{align}
for inter-frame, the velocity steering vector $\mathbf v(v) \in \mathbb C ^ {N_d}$ where
\begin{align}
\label{eq:velocitysteer}
[\mathbf v(v)]_m &\triangleq \exp{\left[-j2\pi f_0 \frac{2v}c mNT_r \right]}.
\end{align}
Additionally, define the vector of ``distortion terms"  $\mathbf c(v) \in \mathbb C^{N N_d}$ where
\begin{align}
[\mathbf c(v)]_{n + mN} \triangleq \exp{\left[-j2\pi n \Delta f \frac{2v}c mNT_r \right]}.
\end{align}
Now let
\begin{align}
\label{eq:phidef}
\boldsymbol \phi(\boldsymbol\theta,\tau,v) \triangleq \mathbf h(\boldsymbol\theta) \otimes \left[\left(\mathbf v(v) \otimes \mathbf r(\tau,v)\right) \odot \mathbf c(v)\right] \in \mathbb C ^{N_{T}N_{R}NN_d},
\end{align}
where $\odot$ is the Hadamard product.
Hence the radar signal component can be expressed in vector form as
\begin{align}
\label{eq:y_R}
\mathbf y_R(k) = \sum_{i\in \mathcal{I}_k} x_i\boldsymbol \phi(\boldsymbol\theta_i,\overline\tau_i(k),v_i).
\end{align}
where the coordinate 
\begin{align}
\overline\tau_i(k) \triangleq \tau_i - t_k \in \left[-\frac T2,\frac T2\right]
\end{align}
is the scatterer's \emph{offset} from the center of the $k$th range cell.

\subsubsection{Communication signal component}
The interference component in the projector output for receiver $q$ is
\begin{align}
y_C^q(m,n,p,k) = \langle s_c(t), s^p(m,n,t - t_k)\rangle.
\end{align}
The power spectral density of $y_C^q$ for any $q$ is
\begin{align}
\mathcal S_C(f) 
&= \sum_{i \in \mathcal C_n}G_{i}(f - f_{i}^C) \left|U^p(f - f_n)\right|^2,
\end{align}
where 
\begin{align}
\mathcal C_n \triangleq \{i \ | \ |f_n - f_i^C| \leq \frac {\Delta f}2 + \frac {B_i}2\}
\end{align}
is the set of indexes of the carriers that overlap with radar pulse $n$. Any communication carrier spectrally overlaps with at least one radar pulse; but in general a particular radar pulse may or may not overlap with any carriers, in which case $\mathcal C_n$ would be empty. We have
\begin{align}
\label{eq:gencommvar}
\E[\left|y_C^q(m,n,p,k)\right|^2] &=  \int_{-\infty}^{\infty} \sum_{i \in \mathcal C_n} G_{i}(f - f_{i}^C) \left|U^p(f - f_n)\right|^2 df,
\end{align}
implying that only the carriers $\mathcal C_n$ may interfere with the radar. Moreover, only a subset of the carriers $\mathcal C_n$ actually interfere because $G_i$ implicitly depends on whether carrier $i$ is in use. Therefore, $\E[\left|y_C^q(m,n,p,k)\right|^2]=0$ whenever 1) $\mathcal C_n=\emptyset$, or 2) none of the carriers $\mathcal C_n$ are in use.

Define $\mathbf B(k) \in \mathbb C ^ {N_T \times N_R \times N_d \times N}$ whose $(p,q,m,n)$ element $B_{pqmn}(k) \triangleq y_C^q(m,n,p,k)$ and let $\mathbf b(k) \triangleq \text{vec}(\mathbf B(k))\in \mathbb C ^ {N_T N_RN_d N }$, such that the $i$th element of $\mathbf b(k)$ is consistent with element $i$ of $\mathbf y_R(k)$. Then the number of nonzero entries in $\mathbf b(k)$ is equal to $N_TN_RN_d$ times the number of occurences of spectral overlap. Intuitively, if the probability of spectrum overlap with an active carrier is small, then $\mathbf b(k)$ will be sparse---a plausible instance of this is explored in Section \ref{sec:simulations}. For now, we assume that $\mathbf b(k)$ has a majority of zeros.

Therefore the projection onto range cell $k$ can be written as
\begin{align}
\label{eq:yfinal}
\begin{split}
\mathbf y(k) = \sum_{i\in \mathcal{I}_k} x_i\boldsymbol \phi(\boldsymbol\theta_i,\overline\tau_i(k),v_i) + \mathbf b(k) + \mathbf e(k),
\end{split}
\end{align}	
where $\mathbf e(k) \sim \mathcal {CN}(0,\sigma^2\mathbf I)$.

\subsection{Optimization Problem Formulation}

The task is to recover the angle-range-doppler image from the radar measurements (\ref{eq:yfinal}). To this end, we construct an on-grid radar model and formulate an optimization problem to jointly recover the image and the interference signal. The following approach images the contents of a single coarse range cell $k$---in practice, the following would be applied to each desired cell. 

The radar data consists of a coherent batch of echo returns due to $N_d$ sweeps, modeled by (\ref{eq:f(t)}). First, the projection operation in (\ref{eq:projections}) isolates the returns due to the scatterers located in range cell $k$, yielding a measurement vector of length $N_T N_RN_d N$, given by (\ref{eq:yfinal}). Next, we assume the scatterers' coordinates in angle-range-velocity space lie on the grid $\mathcal G \subset \mathbb R^4$, where $|\mathcal G| \triangleq M \gg N_T N_RN_d N$. Define $\mathbf \Phi \in \mathbb{C}^{N_T N_RN_d N \times M}$ whose columns form the dictionary $\mathcal D \triangleq \{\boldsymbol\phi(\boldsymbol\theta,\overline\tau,v) \ | \ (\boldsymbol\theta,\overline\tau,v) \in \mathcal G\}$, where $\boldsymbol \phi$ is given by (\ref{eq:phidef}). By the on-grid assumption, we have $\{\boldsymbol \phi(\boldsymbol\theta_i,\overline\tau_i(k),v_i) \ | \ i \in \mathcal I_k\} \subseteq \mathcal D$. Thus, the radar signal component (\ref{eq:y_R}) can be expressed as
\begin{align}
\label{eq:y_R-matvec}
\mathbf y_R(k) = \mathbf \Phi \mathbf w(k),
\end{align}
where $\mathbf w(k) \in \mathbb C^M$ is the vectorized angle-range-doppler image. The nonzero entries of $\mathbf w(k)$ form $\{x_i \ | \ i\in\mathcal{I}_k\}$ and are positioned such that $x_i$ weights $\boldsymbol\phi(\boldsymbol\theta_i,\overline\tau_i(k),v_i)$. Plugging (\ref{eq:y_R-matvec}) into (\ref{eq:yfinal}), we obtain
\begin{align}
\label{eq:linmodel}
\mathbf y = \mathbf \Phi \mathbf w + \mathbf b + \mathbf e,
\end{align}
with the dependence on $k$ hereafter implied.

Sparsity manifests in two forms: $\mathbf b$ is sparse because of the frequency-domain sparsity of the communication signals; $\mathbf w$ is sparse if the radar scene is sparse. Accounting for these properties, we formulate the following optimization problem to \emph{jointly} recover $\mathbf w$ and $\mathbf b$:
\begin{equation}
\begin{aligned}
\label{eq:problem}
& \underset{\mathbf w, \mathbf b}{\text{min}}
& &  \|\mathbf y - \mathbf \Phi \mathbf w - \mathbf b\|_2^2 + \lambda_1 \|\mathbf w\|_1 + \lambda_2\|\mathbf b\|_1.
\end{aligned}
\end{equation}
Given the measurement $\mathbf y$, (\ref{eq:problem}) seeks sparse $\mathbf w$ and $\mathbf b$ that fit (\ref{eq:linmodel}), where the hyperparameters $\lambda_1, \lambda_2 > 0$ control the sparsities. The optimal $\mathbf w$ is the recovered image.

\section{Direct Solver based on ADMM Algorithm}
\label{sec:admmalgo}
	We herein derive the ADMM equations for (\ref{eq:problem}). ADMM is well-suited to handle high-dimensional problems where the objective can be expressed as the sum of convex functions \cite{boyd2011distributed}---as typically is the case in signal processing and machine learning, where dimensionality and regularization terms abound. The problem is split into smaller subproblems which often admit closed-form solutions, so an iteration typically requires only a few matrix-vector multiplies \cite{boyd2011distributed}.
	
	ADMM is often viewed as an approximation of the augmented Lagrange multiplier (ALM) algorithm. ALM solves via gradient ascent the dual of an $\ell_2$-regularized version of the primal problem. Evaluating the dual function entails a joint minimization, which may be prohibitive, so ADMM instead ``approximates" the dual, employing its namesake strategy of minimizing over the variables in an alternating fashion. However the resemblance to ALM is somewhat superficial since each method can be equated to the repeated application of a unique monotone operator, revealing that each method's convergence guarantee is fundamentally different from the other's \cite{eckstein2012ADMM}. Indeed, both methods belong to the broader class of proximal algorithms \cite{parikh2014proximal}. Nonetheless, we derive ADMM via the augmented Lagrangian.

Let $\mathbf A = [\mathbf \Phi \ I_N] \in \mathbb{C}^{N \times (M+N)}$, $\mathbf D_1 = [I_M \ 0]\in \mathbb{R}^{M \times (M+N)}$, $\mathbf D_2 = [0 \ I_N]\in \mathbb{R}^{N \times (M+N)}$, $\mathbf x = \begin{bmatrix}
\mathbf w^T & \mathbf b^T
\end{bmatrix}^T \in \mathbb{C}^{M+N}$, where $I_n$ denotes the $n \times n$ identity matrix. Then (\ref{eq:problem}) is equivalent to
\begin{equation} \label{eq:problem2}
\begin{aligned}
& \underset{\mathbf x}{\text{min}}
& & \|\mathbf y - \mathbf A \mathbf x\|_2^2 + \lambda_1 \| \mathbf D_1 \mathbf x\|_1 + \lambda_2\|\mathbf D_2 \mathbf x\|_1.
\end{aligned}
\end{equation}
We reformulate (\ref{eq:problem2}) as the constrained problem
\begin{equation} \label{OPT:1}
\begin{aligned}
& \underset{\mathbf x, \mathbf z}{\text{min}}
& & \|\mathbf y - \mathbf A \mathbf x\|_2^2 + \lambda_1 \| \mathbf D_1 \mathbf z\|_1 + \lambda_2\|\mathbf D_2 \mathbf z\|_1 \\
& \text{s.t.} 
& & \mathbf x - \mathbf z = 0,
\end{aligned}
\end{equation}
whose augmented Lagrangian is
\begin{align}
L_\rho(\mathbf x, \mathbf z, \mathbf u) 
&= \|\mathbf y - \mathbf A \mathbf x\|_2^2 + \lambda_1 \| \mathbf D_1 \mathbf z\|_1 + \lambda_2\|\mathbf D_2 \mathbf z\|_1 \\ & \nonumber + \frac \rho 2 \|\mathbf x - \mathbf z + \mathbf u\|_2^2 - \frac \rho 2 \|\mathbf u\|_2^2,
\end{align}
where $\mathbf u$ is the scaled dual variable \cite{boyd2011distributed} and $\rho \in \mathbb R$ is a parameter. ALM entails computing the dual function $\underset{\mathbf x, \mathbf z}{\min}\{L_\rho\}$ exactly by jointly minimizing $L_\rho$ over $\mathbf x$ and $\mathbf z$, which may be costly because of the nonlinear term involving $\mathbf x$ and $\mathbf z$. ADMM instead minimizes along the $\mathbf x$ and $\mathbf z$ directions in an alternating fashion. 

``Vanilla" ADMM comprises three steps: minimization of $L_\rho$ over $\bf x$; minimization of $L_\rho$ over $\bf z$; and finally a gradient ascent iteration, incrementing $\mathbf u$ using the gradient w.r.t. $\mathbf u$ of the ``approximate" dual function $\underset{\mathbf z}{\min} \,\underset{\mathbf x}{\min} \, L_\rho(\mathbf x, \mathbf z, \mathbf u)$. Namely, ADMM sequentially solves
\begin{align}
\label{eq:xupdateprob}
\mathbf{x}^{k+1} &= \underset{\mathbf x}{\text{argmin}} \left( \|\mathbf y - \mathbf A \mathbf x\|_2^2 + \frac \rho 2 \|\mathbf x - \mathbf z ^k + \mathbf u^k\|_2^2\right) \\
\label{eq:zupdateprob}
\mathbf z ^{k+1} &=\underset{\mathbf z}{\text{argmin}} \Big( \lambda_1 \| \mathbf D_1 \mathbf z\|_1 + \lambda_2\|\mathbf D_2 \mathbf z\|_1 \\ \nonumber & \hspace{85pt}+ \frac \rho 2 \|\mathbf x^{k+1} - \mathbf z + \mathbf u^k\|_2^2 \Big) \\
\mathbf u ^{k+1} &= \mathbf u ^k + \nabla_\mathbf u L_\rho(\mathbf x^{k+1}, \mathbf z^{k+1}, \mathbf u).
\end{align}
Equation (\ref{eq:xupdateprob}) is an $\ell_2$-regularized least-squares problem, while (\ref{eq:zupdateprob}) can be separated into
\begin{align}
\label{eq:zpartition1}
\mathbf z_1 ^{k+1} &=\underset{\mathbf z_1}{\text{argmin}} \left( \lambda_1\|\mathbf z_1\|_1 + \frac \rho 2 \|\mathbf x_1^{k+1} - \mathbf z_1 + \mathbf u_1^k\|_2^2 \right) \\ 
\label{eq:zpartition2}
\mathbf z_2 ^{k+1} &=\underset{\mathbf z_2}{\text{argmin}} \left(\lambda_2\|\mathbf z_2\|_1 + \frac \rho 2 \|\mathbf x_2^{k+1} - \mathbf z_2 + \mathbf u_2^k\|_2^2 \right),
\end{align}
where $\mathbf z_i \triangleq \mathbf D_i \mathbf z$, $\mathbf x_i^k \triangleq \mathbf D_i \mathbf x^k$ and $\mathbf u_i^k \triangleq \mathbf D_i \mathbf u^k, \ i=1,2$.
The solutions to (\ref{eq:zpartition1}) and (\ref{eq:zpartition2}) are given by the proximal operator of the $\ell_1$-norm, $S_{\kappa}: \mathbb{C}^n \rightarrow \mathbb{C}^n$, called the soft-thresholding operator. Here $S_{\kappa}$ operates elementwise, so that the $i$th element of the output for input $\mathbf a = [a_1 \ \cdots \ a_n]^T$ is 
\begin{align}
[S_{\kappa}(\mathbf a)]_i = \frac{a_i}{|a_i|}*\max(|a_i| - \kappa,0).
\end{align}
Therefore the vanilla ADMM equations for (\ref{eq:problem}) are
\begin{align}
\mathbf x ^{k+1} &= (\mathbf A ^H \mathbf A + \rho I)^{-1}(\mathbf A^H \mathbf y + \rho (\mathbf z^k - \mathbf u^k)) \\
\mathbf z_1 ^{k+1} &= S_{\lambda_1/\rho}(\mathbf x_1 ^{k+1} + \mathbf u_1 ^k)\\
\mathbf z_2 ^{k+1} &= S_{\lambda_2/\rho}(\mathbf x_2 ^{k+1} + \mathbf u_2 ^k)\\
\mathbf u ^{k+1} &= \mathbf u ^k + \mathbf x ^{k+1} - \begin{bmatrix} \mathbf z_1 ^{k+1} \\ \mathbf z_2 ^{k+1} \end{bmatrix}.
\end{align}

Our proposed ADMM algorithm augments vanilla ADMM in two ways. It is known that inserting a relaxation step between the $\mathbf x$ and $\mathbf z$ updates,
\begin{align}
\boldsymbol \xi^{k+1} &= \alpha \mathbf x^{k+1} + (1-\alpha)\mathbf z^k,
\end{align}
where $\alpha \in [0,2]$ is a parameter, may improve convergence speed \cite{boyd2011distributed}. This step also arises naturally in an alternative ADMM derivation \cite{eckstein2012ADMM}. Additionally, we introduce a parameter $\eta \in \mathbb R$ to control the gradient step-size in the $\mathbf u$-update. Finally, the proposed ADMM iteration for (\ref{eq:problem}) is
\begin{align}
\label{eq:admmupdates1}
\mathbf x ^{k+1} &= (\mathbf A ^H \mathbf A + \rho I)^{-1}(\mathbf A^H \mathbf y + \rho (\mathbf z^k - \mathbf u^k)) \\
\label{eq:admmupdates3}
\boldsymbol \xi^{k+1} &= \alpha \mathbf x^{k+1} + (1-\alpha)\mathbf z^k\\
\label{eq:admmupdates4}
\mathbf z_1 ^{k+1} &= S_{\lambda_1/\rho}(\boldsymbol \xi_1 ^{k+1} + \mathbf u_1 ^k)\\
\label{eq:admmupdates5}
\mathbf z_2 ^{k+1} &= S_{\lambda_2/\rho}(\boldsymbol \xi_2 ^{k+1} + \mathbf u_2 ^k)\\
\label{eq:admmupdates2}
\mathbf u ^{k+1} &= \mathbf u ^k + \eta\left(\boldsymbol \xi ^{k+1} - \begin{bmatrix} \mathbf z_1 ^{k+1} \\ \mathbf z_2 ^{k+1} \end{bmatrix} \right).
\end{align}
\begin{figure*}[t]
\includegraphics[width=\linewidth]{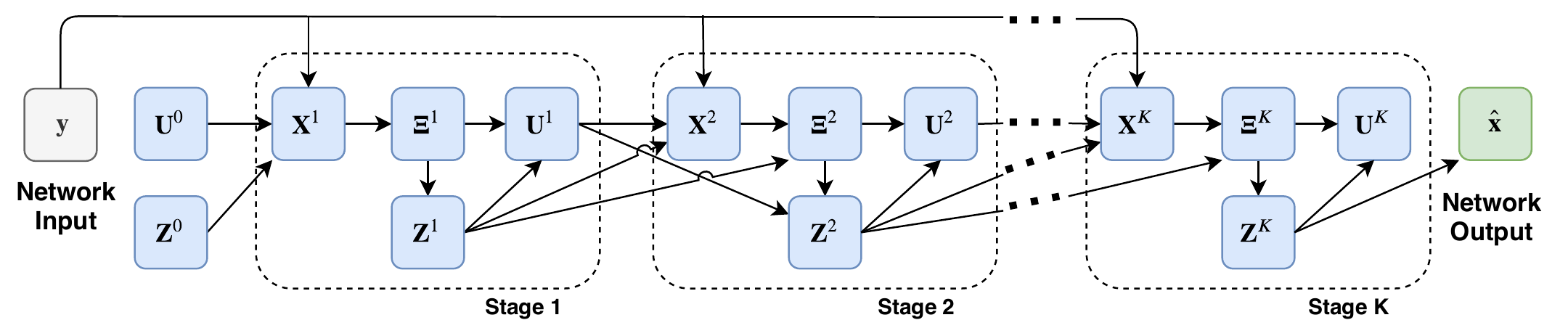}
\caption{Data flow graph for ADMM-net.}
\label{fig:dataflow}
\end{figure*} 
where $\boldsymbol \xi_i^k = \mathbf D_i \boldsymbol \xi^k$, $i=1,\,2$.

The main pitfall of ADMM we aim to address is choosing the parameters, $\{\rho, \alpha, \eta,\lambda_1,\lambda_2\}$ which in general must be tuned for each application. While the basic form of ADMM has a single algorithm parameter $\rho$ and is guaranteed to converge at a linear rate for all $\rho > 0$ \cite{deng2012riceadmm}, in practice the convergence speed as well as accuracy vary significantly with $\rho$. Selection on $\rho$ may be based on the eigenvalues of $\mathbf A$ \cite{teixeira2014admm}. Alternatively, $\rho$ can be updated based on the value of the primal and dual residuals at each iteration \cite{he2000admm}. From our experience, the ADMM parameters primarily influence convergence speed, while the $\ell_1$-regularization parameters affect convergence accuracy. The $\ell_1$ parameters can also be updated at each iteration, e.g. LARS determines a parameter schedule by calculating the solution path for every positive value of the regularization parameter \cite{efron2004lars}. Otherwise, cross-validation can be effective.

The deep unfolding method we present next can be seen as a way of automating hyperparameter cross-validation, wherein algorithm hyperparameters become decision variables for optimizing a measure of algorithm performance.

\section{Complex ADMM-Net}
\label{sec:complexadmmnet}
We herein outline the general unfolded network design process and then detail the proposed ADMM-Net design. Mainstream deep learning software supports only real-valued inputs and parameters, while radar measurements are typically complex-valued, so we have to translate ADMM's complex-valued operations into an equivalent sequence of real-valued operations. Upon network initialization, the network's forward pass is identical to executing a number of ADMM iterations.

\subsection{Towards ADMM-Net}
A neural network is essentially a composition of parameterized linear and nonlinear functions called \emph{layers}, and deep learning is the process of adjusting the layer parameters such that the network emulates some desired mapping. This amounts to optimizing a loss metric quantifying the accuracy of the network's output measured against training data, a putative sample of the desired mapping's input/output behavior. Typically a gradient-based algorithm is used for optimization, and since standard deep learning software packages, such as Tensorflow and PyTorch, employ automatic differentiation to compute gradients, many iterative algorithms can readily be parameterized, cast as a series of network layers, and then optimized as such.

Unfolding an algorithm iteration into a set of feed-forward neural network layers requires specification of a) the functional dependencies between the algorithm iterates and b) the parameters to be learned. Consulting the algorithm's corresponding \emph{data flow graph} aids the design process. Fig. \ref{fig:dataflow} depicts the data flow graph for the proposed ADMM iteration. Each node corresponds to an iterate, and an arrow indicates functional dependence between two iterates. The iterate associated with an arrow's head is a function of the iterate associated with the tail. The neural network receives one layer per node, such that the inputs to the layer associated with a node $v$ are the tails of all arrows directed to $v$. A layer's input/output mapping is defined based on the corresponding iterate's update equation in the original algorithm, or a generalized version thereof. Therefore if the algorithm comprises $n$ update equations, every $n$ consecutive layers of the unfolded network correspond to a single algorithm iteration---we refer to this as a network ``stage" \cite{yang2016admmnet}; see the nodes enclosed by the dashed-lines in Fig. \ref{fig:dataflow}. 

\subsection{ADMM-Net Structure}
ADMM-Net has layer operations based on (\ref{eq:admmupdates1})-(\ref{eq:admmupdates2}). Stage $k$ of the network consists of a reconstruction layer $\mathbf X^k$ that corresponds to the $\mathbf x$-update, a relaxation layer $\mathbf \Xi^k$ that corresponds to the $\boldsymbol \xi$-update, a nonlinear transform layer $\mathbf Z^k$ that corresponds to the $\mathbf z$-update, and a dual update layer $\mathbf U^k$ that corresponds to the $\mathbf u$-update. In addition to learning the ADMM algorithm parameters in each layer, we also parameterize the linear transformations in the $\mathbf x$-update, initializing them as prescribed by ADMM. 

\textbf {Network Input}: The network input $\mathbf y \in \mathbb C^N$ enters the network via the reconstruction layers. 

\begin{figure*}
\centering
	\subfloat[][]{\includegraphics[width=62mm]{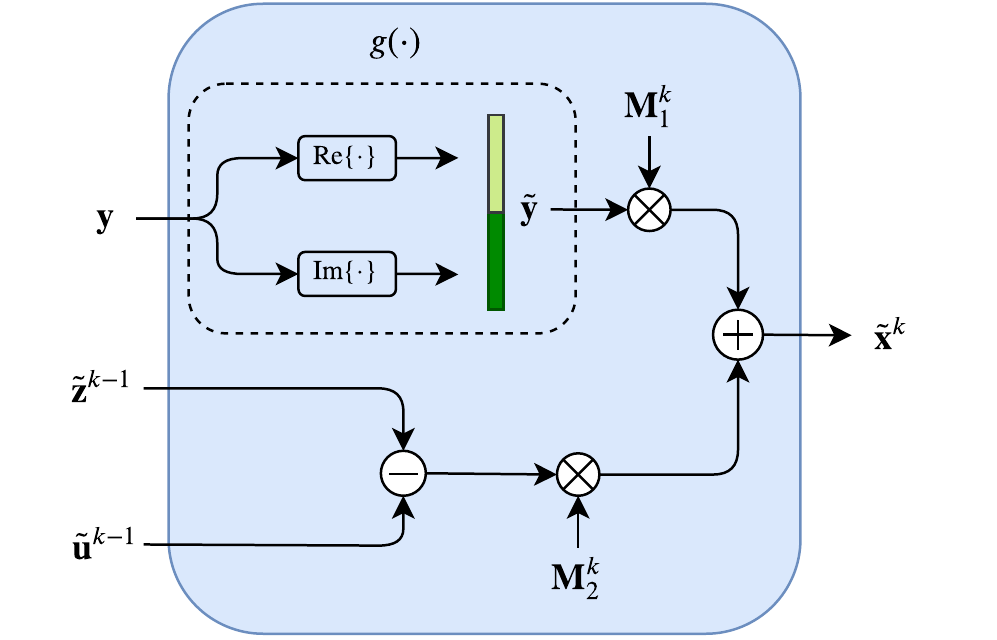}}
	\subfloat[][]{\includegraphics[width=97mm]{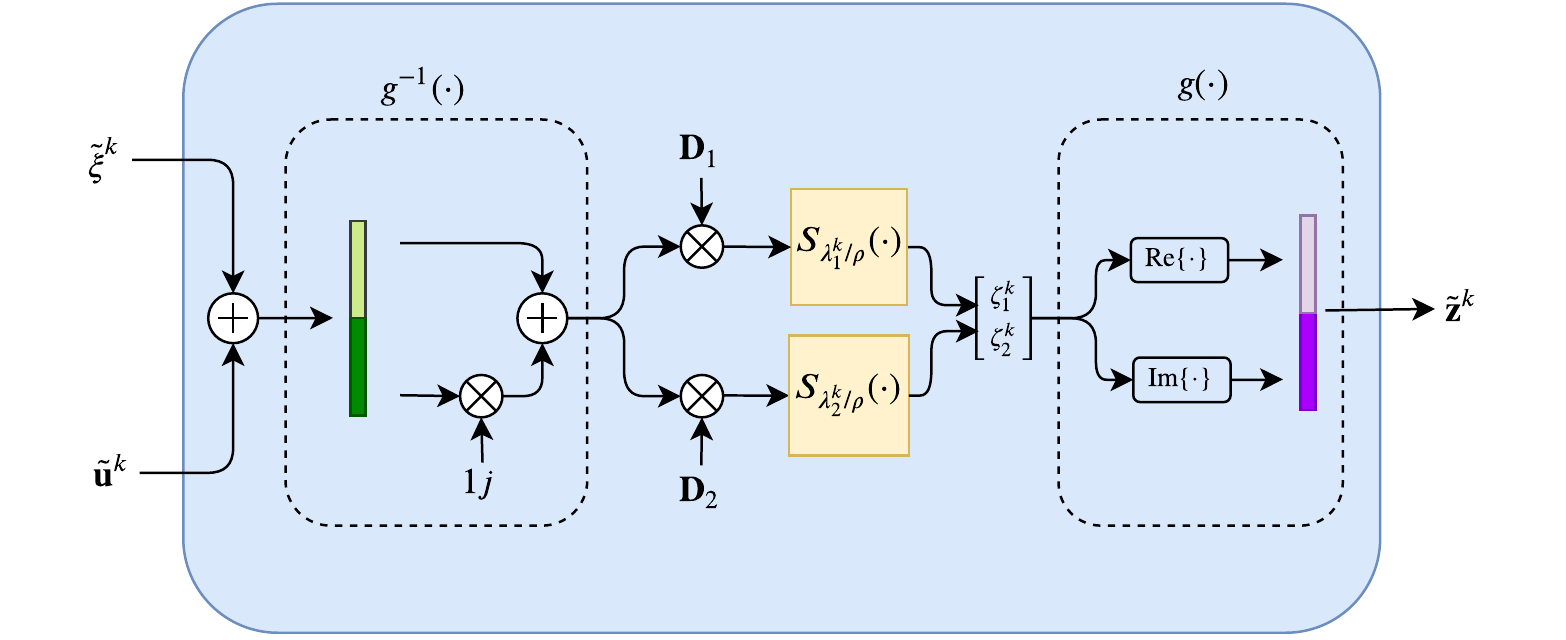}}
	\caption{Block diagrams for reconstruction (a) and nonlinear transform (b) layers.}
	\label{fig:blockdiagrams}
\end{figure*}

\textbf {Reconstruction Layer}: This layer performs the complex $\mathbf x$-update prescribed by ADMM. The inputs to this layer are the network input ${\mathbf y}\in \mathbb C^N$, and $\tilde{\mathbf z}^{k-1},\tilde{\mathbf u}^{k-1} \in \mathbb R ^ {2M}$ from stage $k-1$.
The output of the stage $k$ reconstruction layer is
\begin{align}
\label{reconlayer}
\tilde{\mathbf x} ^{k} &= \mathbf M_1^{k} g(\mathbf y) + \mathbf M_2^{k} (\tilde{\mathbf z}^{k-1} - \tilde{\mathbf u}^{k-1}),
\end{align}
and hence $\tilde{\mathbf x}^k \in \mathbb R ^ {2M}$. The entries of the matrices $\mathbf M_1^{k} \in \mathbb R ^ {2M\times2N}$ and $\mathbf M_2^{k} \in \mathbb R ^ {2M\times2M}$ are learnable parameters. The function $g:\mathbb C^{N}\rightarrow\mathbb R^{2N}$ vertically concatenates the input's real and imaginary parts into a single real-valued vector: if ${\mathbf y} \in \mathbb C^{N}$, then
\begin{align}
g(\mathbf y) := \begin{bmatrix} \real{\mathbf y} \\ \imag{\mathbf y} \end{bmatrix} \in \mathbb R^{2N}.
\end{align} 
The block diagram for $g$ is shown in Fig.~\ref{fig:blockdiagrams}(a). Thus $\tilde{\mathbf x}^k$ corresponds to ``stacking" the real and imaginary parts of (\ref{eq:admmupdates1}) into a single vector, i.e. $\tilde{\mathbf x}^k \equiv \begin{bmatrix} \mathrm{Re}\{{{\mathbf x} ^{k}}^T\} & \mathrm{Im}\{{{\mathbf x} ^{k}}^T\} \end{bmatrix}^T$. The values $\tilde{\mathbf z}^0=0$ and $\tilde{\mathbf u}^0=0$ are used for the first reconstruction layer.

Fig.~\ref{fig:blockdiagrams}(a) illustrates the $k$th reconstruction layer: the real and imaginary parts of the complex-valued observation $\mathbf y$ are vertically concatenated via $g$ to form $\tilde{\mathbf y}$; $\mathbf M_1^{k}$ premultiplies $\tilde{\mathbf y}$ and $\mathbf M_2^{k}$ premultiplies $\tilde{\mathbf z}^{k-1} - \tilde{\mathbf u}^{k-1}$; the two resulting vectors are summed to obtain the layer output $\tilde{\mathbf x}^{k}$.

\textbf {Relaxation Layer} (stage $k$): The output of this layer is
\begin{align}
\label{eq:relaxaxtionlayer}
\tilde{\boldsymbol \xi}^{k} &= \alpha^k\tilde{\mathbf x} ^{k} + (1-\alpha^k)\tilde{\mathbf z} ^{k-1},
\end{align}
where $\alpha^k > 0$ is a learnable parameter. The output $\tilde{\boldsymbol \xi}^{k} \in \mathbb R^{2M}$ is the concatenation of the real and imaginary parts of (\ref{eq:admmupdates3}), i.e. $\tilde{\boldsymbol \xi}^k \equiv \begin{bmatrix} \mathrm{Re}\{{{\boldsymbol \xi} ^{k}}^T\} & \mathrm{Im}\{{{\boldsymbol \xi} ^{k}}^T\} \end{bmatrix}^T$.

\textbf {Nonlinear Transform Layer}: This layer applies the soft-thresholding operation as in the ADMM $\mathbf z$-update (\ref{eq:admmupdates4})-(\ref{eq:admmupdates5}). The output of the $K$th nonlinear transform layer is sent to the network output layer. The layer output is given by
\begin{align}
\boldsymbol \zeta_1^k &= S_{\lambda_1^{k}}({\mathbf D}_1 g^{-1}(\tilde{\boldsymbol \xi} ^{k} + \tilde{\mathbf u} ^{k-1})) \\
\boldsymbol \zeta_2^k& = S_{\lambda_2^{k}}({\mathbf D}_2 g^{-1}(\tilde{\boldsymbol \xi} ^{k} + \tilde{\mathbf u} ^{k-1})) \\
\tilde{\mathbf z} ^{k} &=
g \left( \begin{bmatrix}
\boldsymbol \zeta_1^k \\
\boldsymbol \zeta_2^k
\end{bmatrix} \right),
\end{align}
where $\lambda_1^k, \lambda_2^k > 0$ are the learnable $\ell_1$-regularization parameters and $S_{\kappa}(\mathbf a)_i = \frac{a_i}{|a_i|}*\max(|a_i| - \kappa,0)$ is the soft-thresholding operator. The operation $g^{-1}:\mathbb R^{2M}\rightarrow\mathbb C^{M}$ forms a complex vector out of the top and bottom halves of the input vector: if $\tilde{\mathbf x} \in \mathbb R^{2M}$ then
\begin{align}
\label{eq:ginv}
g^{-1}(\tilde{\mathbf x}) := \tilde{\mathbf x}[0:M-1] + j \tilde{\mathbf x}[M:2M-1] \in \mathbb C^{M},
\end{align}
where the notation $\mathbf a[k:l]$ refers to a vector containing the $k$th through the $l$th components inclusive, of the vector $\mathbf a$. The block diagram for $g^{-1}$ is shown on the left-hand side of Fig.~\ref{fig:blockdiagrams}(b). The matrices $\mathbf D_1 = [I_M \ 0]\in \mathbb{R}^{M \times (M+N)}$ and $\mathbf D_2 = [0 \ I_N]\in \mathbb{R}^{N \times (M+N)}$ partition $\mathbf z$ as in (\ref{eq:zpartition1})-(\ref{eq:zpartition2}). 

Fig.~\ref{fig:blockdiagrams}(b) illustrates this layer's operations: the layer inputs are summed and input to $g^{-1}$; the output is partitioned via premultiplication by ${\mathbf D}_1$ and ${\mathbf D}_2$; soft-thresholding is applied with the respective thresholding parameters $\lambda_1$ and $\lambda_2$; the outputs are concatenated into $\begin{bmatrix}
{\boldsymbol \zeta_1^k} ^T&
{\boldsymbol \zeta_2^k}^T
\end{bmatrix}^T,$ whose real and imaginary parts are subsequently concatenated into the real-valued vector via $g$, yielding the output $\tilde{\mathbf z}^k$.

\textbf {Dual Update Layer}: The output of this layer is
\begin{align}
\label{eq:dualupdatelayer}
\tilde{\mathbf u} ^{k} &= \tilde{\mathbf u} ^{k-1} + \eta^{k} ( \tilde{\boldsymbol \xi} ^{k} - \tilde{\mathbf z} ^{k} ),
\end{align}
where $\eta^k$ is a learnable parameter corresponding to the gradient step size. The variable $\tilde{\mathbf u}^{k} \in \mathbb R^{2N}$ corresponds to the concatenation of the real and imaginary parts of (\ref{eq:admmupdates3}), i.e. $\tilde{\mathbf u}^k \equiv \begin{bmatrix} \mathrm{Re}\{{{\mathbf u} ^{k}}^T\} & \mathrm{Im}\{{{\mathbf u} ^{k}}^T\} \end{bmatrix}^T$. 

\textbf{Network Output}: The network output is derived from the output of the final nonlinear transform layer $\tilde{\mathbf z}^K$ via
\begin{align}
\hat{\mathbf x} = g^{-1}(\tilde{\mathbf z}^K),
\end{align} where $g^{-1}$ is defined in (\ref{eq:ginv}).

\subsection{Training Details}
 
\subsubsection{Parameter set}

Stage $k$ of the network has learnable parameters $\{\mathbf M_1^k, \mathbf M_2^k, \alpha^k, \lambda_1^k, \lambda_2^k, \eta^k\}$. The scalar parameters are initialized according to either theoretically or empirically justified values, as detailed in Section \ref{sec:simulations}. The matrices $\mathbf M_1^{k} \in \mathbb R ^ {2M\times2N}$ and $\mathbf M_2^{k} \in \mathbb R ^ {2M\times2M}$ are initialized such that the reconstruction layer's operation is initially equivalent to (\ref{eq:admmupdates1}). All stages are identically initialized according to
\begin{align}
\label{eq:M1M2init}
\mathbf M_1^k &\leftarrow \begin{bmatrix}
\mathrm{Re}\{\mathbf P\mathbf A^T\} & -\mathrm{Im}\{\mathbf P\mathbf A^T\} \\
\mathrm{Im}\{\mathbf P\mathbf A^T\} & \mathrm{Re}\{\mathbf P\mathbf A^T\}
\end{bmatrix} \in \mathbb R ^ {2M\times2N} \\ \nonumber
\mathbf M_2^k &\leftarrow \begin{bmatrix}
\mathrm{Re}\{\rho\mathbf P\} & -\mathrm{Im}\{\rho\mathbf P\} \\
\mathrm{Im}\{\rho\mathbf P\} & \mathrm{Re}\{\rho\mathbf P\}
\end{bmatrix}\in \mathbb R ^ {2M\times2M},
\end{align}
where $\mathbf P \triangleq (\mathbf A ^T \mathbf A + \rho I)^{-1}$.

\subsubsection{Training data generation}
\label{sec:trainingdatagen}
Training data pairs $\{({\mathbf x}_i, {\mathbf y}_i)\}_{i=1}^{N_{\mathrm{train}}}$ are generated as follows. The complex-valued ground truth ${\mathbf x}_i = \begin{bmatrix} {\mathbf w}_i^T & {\mathbf b}_i^T \end{bmatrix}^T \in \mathbb C^M$ is created such that $\mathbf x_i$ and $\mathbf b_i$ satisfy desired sparsity levels, where the nonzero elements are sampled from a distribution dictated by the physical model. The complex-valued measurements are then generated by ${\mathbf y}_i = \mathbf A {\mathbf x}_i + \mathbf e_i$ where $\mathbf e_i \sim \mathcal N(0,\sigma^2\mathbf I)$ with noise level $\sigma$.

\subsubsection{Loss function}

The loss function of the network is the mean-squared error 
\begin{align}
\mathcal L(\hat {\mathbf x}_i ; \mathbf x_i) = \frac 1 {N_{\mathrm{train}}}\sum_{i=1}^{N_{\mathrm{train}}} \|\mathbf x_i - \hat {\mathbf x}_i\|_2^2,
\end{align} 
where $\hat {\mathbf x}$ is the network's output and $\mathbf x_i$ is the $i$th training sample. 

\section{Simulations}
\label{sec:simulations}
We compare the performance ADMM-net, ADMM, and the CVX semi-definite program solver in angle-range-velocity imaging in a simulated interference environment where a MIMO stepped-frequency radar shares spectrum with the SC-FDMA system introduced in Section \ref{sec:commsignal}, and further specified in Section \ref{subsec:SC-FDMA}. 

\subsection{Angle-range-velocity imaging} 
Simulated radar measurements are generated with the on-grid model (\ref{eq:linmodel}). The simulated (toy-sized) stepped-frequency MIMO radar parameters are listed in Table \ref{tab:radarparams}. The scattering coefficients $x_i$ are independently sampled from $\mathcal{CN}(0,\sigma_x^2)$, where $\sigma_x^2$ is the variance.  The columns of $\mathbf \Phi$ are scaled to have unit norm. Without loss of generality, we consider the radar processing for the range cell $k=0$. 

The Tx and Rx arrays are co-planar uniform linear arrays, with respective normalized element spacings $d_T$ and $d_R$ (normalized by the start carrier wavelength $f_0 /c$), arranged in a cross-shaped geometry \cite{ungan2020millscross}. The array response to a scatterer at angular coordinates $(\theta_{1}, \theta_{2}) \in \mathbb R^2$, where $\theta_{1}$ is the direction relative to the Rx array and $\theta_{2}$ is the direction relative to the Tx array, is given by
\begin{align}
\mathbf h(\theta_1,\theta_2) = \mathbf h_R(\theta_1) \otimes \mathbf h_T(\theta_2),
\end{align}
where
\begin{align}
\mathbf h_R(\theta_1) &\triangleq \begin{bmatrix}
1 & e^{-j2\pi d_R\theta_1} & \cdots & e^{-j2\pi d_R\theta_1 (N_R-1)}
\end{bmatrix} \\
\mathbf h_T(\theta_2) &\triangleq \begin{bmatrix}
1 & e^{-j2\pi d_T\theta_2} & \cdots & e^{-j2\pi d_T\theta_2 (N_T-1)}.
\end{bmatrix}
\end{align}
We let $\mathcal G = \mathcal T \times \mathcal V \times \Theta_1 \times  \Theta_2$, where
\begin{align}
\mathcal T &= \{T m/M_\tau \ | \ m=-M_\tau/2 , \dots, M_\tau/2-1\} \\
\mathcal V &= \{v_{\mathrm{max}}m/M_v \ | \ m=-M_v/2 , \dots, M_v/2-1\} \\
\Theta_1 &= \{m/M_{\theta_1} \ | \ m=0,1,\dots,M_{\theta_1-1}\} \\
 \Theta_2 &= \{m/M_{\theta_2} \ | \ m=0,1,\dots,M_{\theta_2-1}\} 
\end{align} are the delay, velocity, and angle grids, and  $M_\tau$,$M_v$, and $M_{\theta_1}$ and $M_{\theta_2}$ are the respective grid sizes. Recall $\overline\tau \in [-\frac T2,\frac T2]$ is the offset from the center of the coarse range cell; the absolute delay $\tau$ is recovered via $\tau = \overline\tau + t_k$, where $t_k$ is the center of the coarse range cell. 

We choose, $M_\tau = 5$, $M_v = 5$, and $M_{\theta_1} =3, M_{\theta_2} = 2$; hence $\mathbf \Phi \in \mathbb C^{64 \times 150}$. To avoid aliasing, we require $|v| \leq \frac c {4f_0NT_r} \triangleq v_{\mathrm{max}}$, and assuming $d_R=d_T=1$, we require $0\leq{\theta_1}\leq 1$ and $0\leq{\theta_2}\leq 1$. The maximum unambiguous absolute range is thus $R_{\mathrm{max}}=c T_r/2$ meters. Each coarse range cell is of size $cT/2 = 150$ meters; the conventional, DFT-based range profile resolution is $\frac {cT} {2N} = 37.5$ meters. The maximum unambiguous velocity is $\pm v_{\mathrm{max}} = \pm \frac c {4f_0NT_r}$.

\begin{table}[!htb]
\caption{Radar simulation parameters}
\label{tab:radarparams}
\centering
\begin{tabular}{cccccc}  
\toprule
Symbol&Value&Description\\
\midrule
$N$ & 4 & No. frequency steps\\  
$N_d$ & 4 & No. sweeps \\
$N_T$ & 2 & No. transmitters\\
$N_R$ & 2 & No. receivers \\
$f_0$ & 2 GHz & Start frequency \\
$\Delta f$ & 1 MHz & Frequency step size\\
$T$ & 1 $\mu$s & Pulse duration\\
$T_r$ & 66 $\mu$s & Pulse-repetition interval \\
$R_{\mathrm{max}}$ & 9900 m & Max. unambiguous range\\
$v_{\mathrm{max}}$ & $\pm142$ m/s & Max. unambiguous velocity\\
$d_T$ & 1 & Tx array normalized spacing \\
$d_R$ & 1 & Rx array normalized spacing \\
$\sigma$ & various & AWGN variance \\
$\sigma_x^2$ & 1 & Scattering coefficient variance \\
$x_i$ & $\sim\mathcal{CN}(0,\sigma_x^2)$ & Scattering coefficient $i$ \\
\bottomrule
\end{tabular}
\end{table}

\subsection{SC-FDMA}
\label{subsec:SC-FDMA}

Table \ref{tab:scfdmaparams} lists the simulated SC-FDMA system parameters. Without loss of generality, in the simulations we make the following assumptions:
\begin{enumerate}
\item $f_0^C = f_0$. The radar and SC-FDMA system have the same start frequency, $f_0$.
\label{startassumption}
\item $N_cK\Delta f^C = N\Delta f$. The SC-FDMA bandwidth equals the radar sweep bandwidth, and therefore the SC-FDMA system is the only source of interference. The extension to multiple interference sources is straightforward since each source would occupy a distinct frequency band; an analysis along the lines presented here would be carried out for each interference source.
\label{bandassumption}
\item $\frac{\Delta f}{K\Delta f^C} \triangleq L \in \mathbb Z^+$. The sweep bandwidth is an integer multiple of the channel bandwidth. For example, the coherence bandwidth is typically $\sim 0.5 \ \mathrm{MHz}$ and typically $\Delta f \geq 1 \ \mathrm{MHz}$.
\label{sweepassumption}
\item $\gamma_i = \gamma$ for all $i$, where $\gamma>0$ is a constant. 
\item We suppose the scheduling takes place on a PRI-by-PRI basis (in LTE, resource blocks are allocated in time intervals on the order of 1 millisecond, while the radar sweep duration may be tens of milliseconds). Let $\Omega$ denote the sample space of all possible active channel configurations---i.e. the power set of $\{n\in\mathbb Z \ | \ 0\leq n \leq N_c-1\}$---and let $A_i \subset \Omega$ denote the event channel $i$ is active during any given radar pulse, where the probability of $A_i$ is $P(A_i)$. We assume a random sample from $\Omega$ is drawn every PRI. 
\item $\{a_{ik}(n_c):\forall i,k,n_c\}$ are i.i.d., uncorrelated, and normalized, such that
\begin{align}
\E[a_{ik}(n_c)a_{ik'}(n_c)^*|A_i] = \begin{cases}
1 & \mathrm{if} \ k=k'  \\
0 & \mathrm{if} \ k \neq k'
\end{cases}.
\end{align}
In practice, the cyclic prefix violates the uncorrelatedness assumption, but the discrepancy will be small to the extent that the length of the channel impulse response is small relative to the symbol duration (e.g. in LTE the cyclic prefix duration is around $7\%$ of the data symbol duration). Also, if the symbols are normalized, then by the norm-preserving property of the isometric DFT, the original data symbols belong to a normalized symbol set.
\label{aikassump}
\item $a_{ik}(n_c)$ and $h_i$ are mutually independent for all $i$, $k$, and $n_c$.
\label{mutindassumption}
\item  $P(A_i) \triangleq \epsilon \in [0,1]$ for all $i$. This implies $\mathbf b$ is sparse with high probability whenever $\epsilon$ is small. 
\end{enumerate}
 
\begin{table}[!htb]
\caption{SC-FDMA parameters}
\label{tab:scfdmaparams}
\centering
\begin{tabular}{cccccc}  
\toprule
Parameter&Value&Description\\
\midrule
$f_0^C$ & 2 GHz & Start frequency \\
$K\Delta f^C$ & 0.5 MHz & Channel bandwidth\\  
$N_c$ & 8 & Number of channels \\
$\gamma$ & 1 & Power assigned to each channel \\
$\epsilon$ & various & Proportion of active channels \\
$\beta$ & various & Variance of channel fading coefficient \\
\bottomrule
\end{tabular}
\end{table}

\subsection{Signal-to-Noise Ratio}
We define the signal-to-noise ratio (SNR) for a given range cell $k$ as
\begin{align}
\mathrm{SNR} &\triangleq \frac {\mathbb E \left[ \| \mathbf y_R(k)\|_2^2 \right]}{\mathbb E[\| \mathbf e(k) \|_2^2]}
\end{align}
where $\mathbf y_R$ is given in (\ref{eq:y_R}) and $\mathbf e(k) \sim \mathcal {CN}(0,\sigma^2\mathbf I)$.
\subsection{Signal-to-Interference Ratio}
The signal-to-interference ratio (SIR) for a given range cell $k$ is defined as
\begin{align}
\label{eq:SIR}
\mathrm{SIR} &\triangleq \frac {\mathbb E \left[ \| \mathbf y_R(k)\|_2^2 \right]}{\mathbb E[\| \mathbf b(k) \|_2^2]}
\end{align}
where $\mathbf b$ is given by (\ref{eq:yfinal}). 

\subsection{Algorithm Specifications}
\subsubsection{ADMM-Net}
\label{sec:algoparamsADMMnet}
Unless otherwise indicated, training sets were of size $N_{\mathrm{train}} = 4.5\times10^6$ and the networks were trained for 45 epochs, i.e. full passes over the training set. The Adam \cite{kingma2014adam} optimizer was used with parameters $\beta_1 = 0.9$, $\beta_2 = 0.999$ and a batch size of $500$. The Adam learning rate was initialized to $10^{-3}$ and multiplied by $10^{-1}$ every $15$ epochs. All networks were implemented and trained with the Keras API in Tensorflow 2.

The nonzero entries of ${\mathbf w}_i$ were generated i.i.d. $\mathcal{CN}(0,1)$. The nonzero entries of ${\mathbf b}_i$ were generated i.i.d. $\mathcal{CN}(0,\beta)$, where $\beta$ was chosen to satisfy a given $\mathrm{SIR}$. The noise $\mathbf e$ was drawn from $\mathcal{CN}(0,\sigma^2\mathbf I)$, where $\sigma^2$ was chosen to satisfy a given $\mathrm{SNR}$. See Section \ref{sec:trainingdatagen} for more details regarding training data generation.

The scalar network parameters were initialized identically for all layers $k$ as
\begin{align}
\alpha^k &= 1.5 & 
\lambda_1^k &= 0.01 \\
\eta^k &= 1 & 
\lambda_2^k &= 0.005 \\
\rho^k&=0.01.
\end{align}
The value for $\lambda_2^k$ was determined by cross-validation; $\eta^k$ was set to accord with the ``vanilla" ADMM equations; $\alpha^k$ was set as recommended \cite{boyd2011distributed}; $\rho^k$ was set as recommended \cite{fastflexibleadmm}.
The matrices $\mathbf M_1^k$ and $\mathbf M_2^k$ are initialized according to (\ref{eq:M1M2init}) so that they coincide with ADMM.

\subsubsection{ADMM}
An ADMM iteration is given by Eqs. (\ref{eq:admmupdates1})-(\ref{eq:admmupdates2}). We use the following parameter values:
\begin{align}
\alpha &= 1.5 & 
\lambda_1 &= 0.01 \\
\eta &= 1 & 
\lambda_2 &= 0.005 \\
\rho&= 0.01.
\end{align}
The justification for these values is the same as that for the ADMM-net parameter initialization (Section \ref{sec:algoparamsADMMnet}).

\subsubsection{CVX}
For CVX, we used the semi-definite program (SDP) solver on the problem
\begin{equation} \label{OPT:CVXSDP}
\begin{aligned}
& \underset{\mathbf x, \mathbf z}{\text{min}}
& & \|\mathbf z\|_2^2 + \lambda_1 \| \mathbf D_1 \mathbf x\|_1 + \lambda_2\|\mathbf D_2 \mathbf x\|_1 \\
& \text{s.t.} 
& & \mathbf z = \mathbf y - \mathbf A \mathbf x
\end{aligned}
\end{equation}
with parameter values
\begin{align}
\lambda_1^k &= 0.01 & \lambda_2^k &= 0.005,
\end{align}
where $\lambda_2$ was found through cross-validation.

\subsubsection{ADMM Single-Penalty}
To highlight the benefit of the proposed two-penalty objective (\ref{eq:problem}), we also consider the problem
\begin{equation}
\begin{aligned}
\label{eq:problemsinglepenalty}
& \underset{\mathbf w}{\text{min}}
& &  \|\mathbf y - \mathbf \Phi \mathbf w\|_2^2 + \lambda_1 \|\mathbf w\|_1.
\end{aligned}
\end{equation}
We ran the associated ADMM algorithm with parameters
\begin{align}
\alpha &= 1.5 & 
\lambda_1 &= 0.5 \\
\eta &= 1 & 
\rho&= 0.5.
\end{align}

\begin{figure}
\includegraphics[width=75mm]{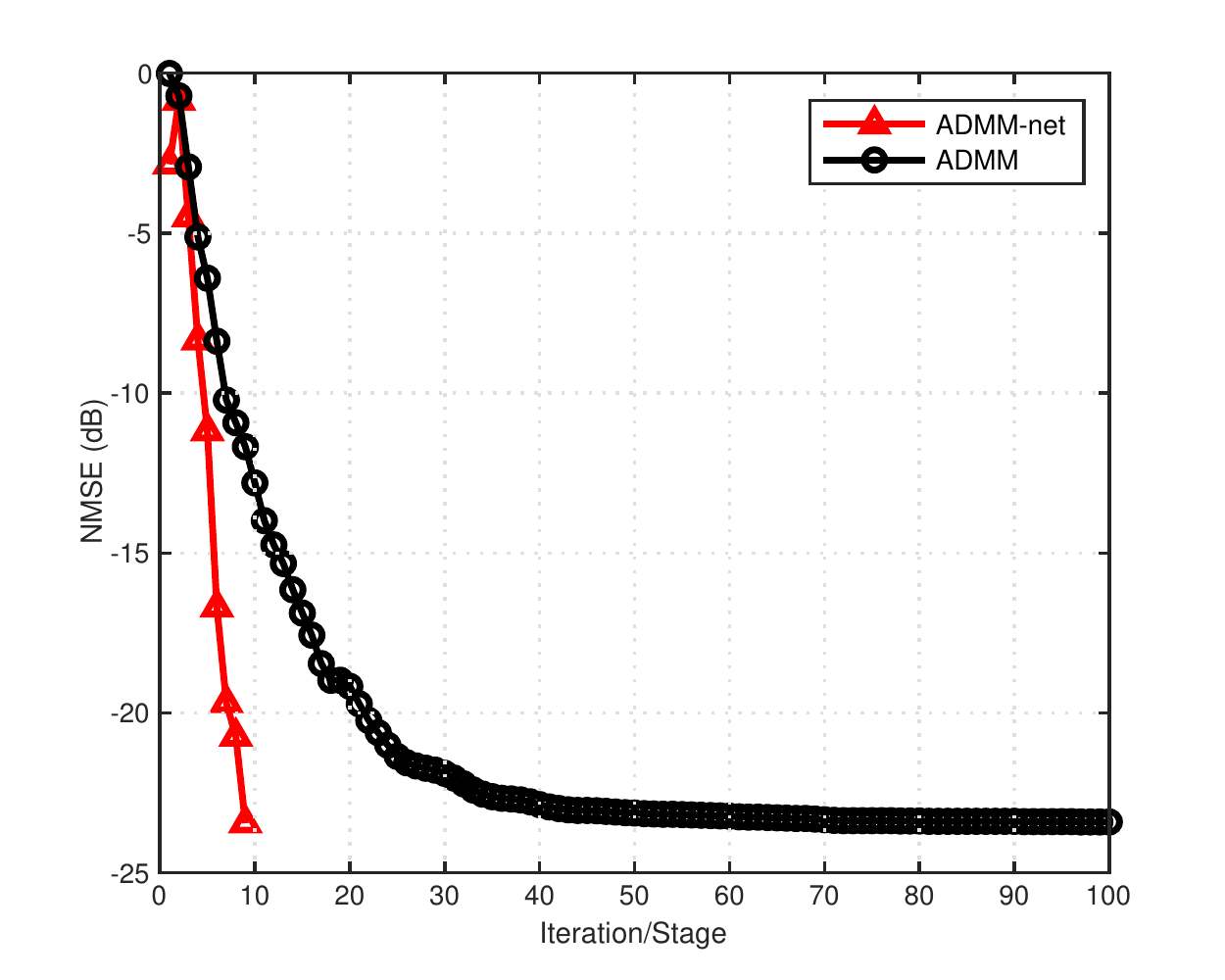}
\caption{$\mathrm{NMSE}$ versus iteration/stages of ADMM/ADMM-Net. The training and test sets have $\mathrm{SNR} = \infty$, $\|\mathbf w\|_0 = 2$, $\|\mathbf b\|_0 = 16$ and $\mathrm{SIR} = 0 \ \mathrm{dB}$.}
\label{fig:siso_num_stages_experiment}
\end{figure}

\subsection{Results}
The experiments probe the network's performance and robustness along four dimensions: network depth (number of stages), $\mathrm{SNR}$, $\mathrm{SIR}$, and sparsity level. We evaluate the candidate methods via the average normalized mean squared error ($\mathrm{NMSE}$) of their estimates, defined as
\begin{align}
\mathrm{NMSE} = 10\log_{10} \left[\frac 1{N_{\mathrm{test}}} \sum_{i=1}^{N_{\mathrm{test}}}\frac {\|\mathbf x_i - \hat {\mathbf x}_i\|_2^2}{\| \mathbf{x}_i \|_2^2}\right] \ \mathrm{dB},
\end{align}
where $\hat{\mathbf x}_i$ is the algorithm output and $\mathbf x_i$ is the ground truth. The same test set, with $N_{\mathrm{test}} = 10^3$, was used to evaluate all algorithms.

For each data set property ($\mathrm{SNR}$, sparsity, etc.) we train several networks, each on a different training set. Each training set contains samples with either a particular value or a random distribution of values for the property. Next, we report the results of each experiment. 

\subsubsection{Network stages}
Fig. \ref{fig:siso_num_stages_experiment} shows algorithm $\mathrm{NMSE}$ (dB) versus the number of stages/iterations for ADMM-Net/ADMM for the case $\mathrm{SNR} = \infty$ (i.e., $\sigma^2=0$), $\|\mathbf w\|_0 = 2$, and $\|\mathbf b\|_0 = 16$. The 9-stage ADMM-Net achieves an error of $-23.45 \ \mathrm{dB}$ while ADMM converges to $-23.48 \ \mathrm{dB}$ in 195 iterations (see (\ref{eq:convergencecriterion}) for the convergence criterion). The ADMM-Net was trained for 45 epochs on $5\times10^6$ samples.

\subsubsection{SNR}
Five networks were trained: four on data sets with deterministic $\mathrm{SNR}$s in $\{5,\, 10,\, 15\}$ and one with random $\mathrm{SNR}$s drawn from $\mathrm{uniform}(5,\,20)$. Fig. \ref{fig:SNR-exp} plots algorithm $\mathrm{NMSE}$ (dB) versus $\mathrm{SNR}$, where in all cases $\|\mathbf w\|_0 = 2$, $\|\mathbf b\|_0 = 16$ (25\% spectral overlap), and $\mathrm{SIR} = 0 \ \mathrm{dB}$. The points on red curve are the $\mathrm{NMSE}$s of the networks trained on data with a deterministic $\mathrm{SNR}$ equal to the point's abscissa; the points on the blue curve are the $\mathrm{NMSE}$s of the single network trained on the random $\mathrm{SNR}$ data.

\subsubsection{SIR}
\begin{figure}[t]
\includegraphics[width=75mm]{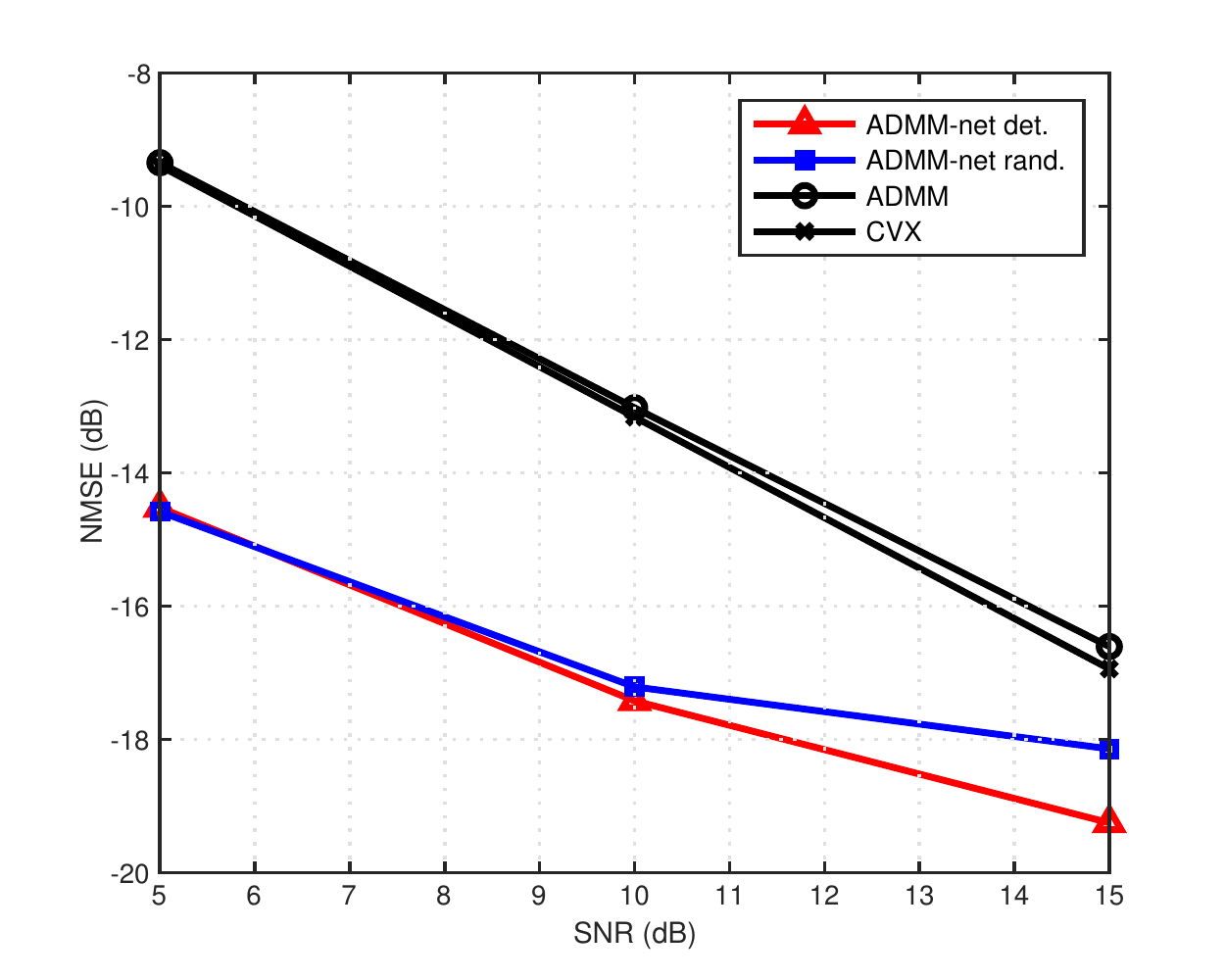}
\caption{$\mathrm{NMSE}$ versus $\mathrm{SNR}$. 2 scatterers, 25\% spectrum overlap, $\mathrm{SIR}=0$.}
\label{fig:SNR-exp}
\end{figure}

\begin{figure}[t]
\includegraphics[width=75mm]{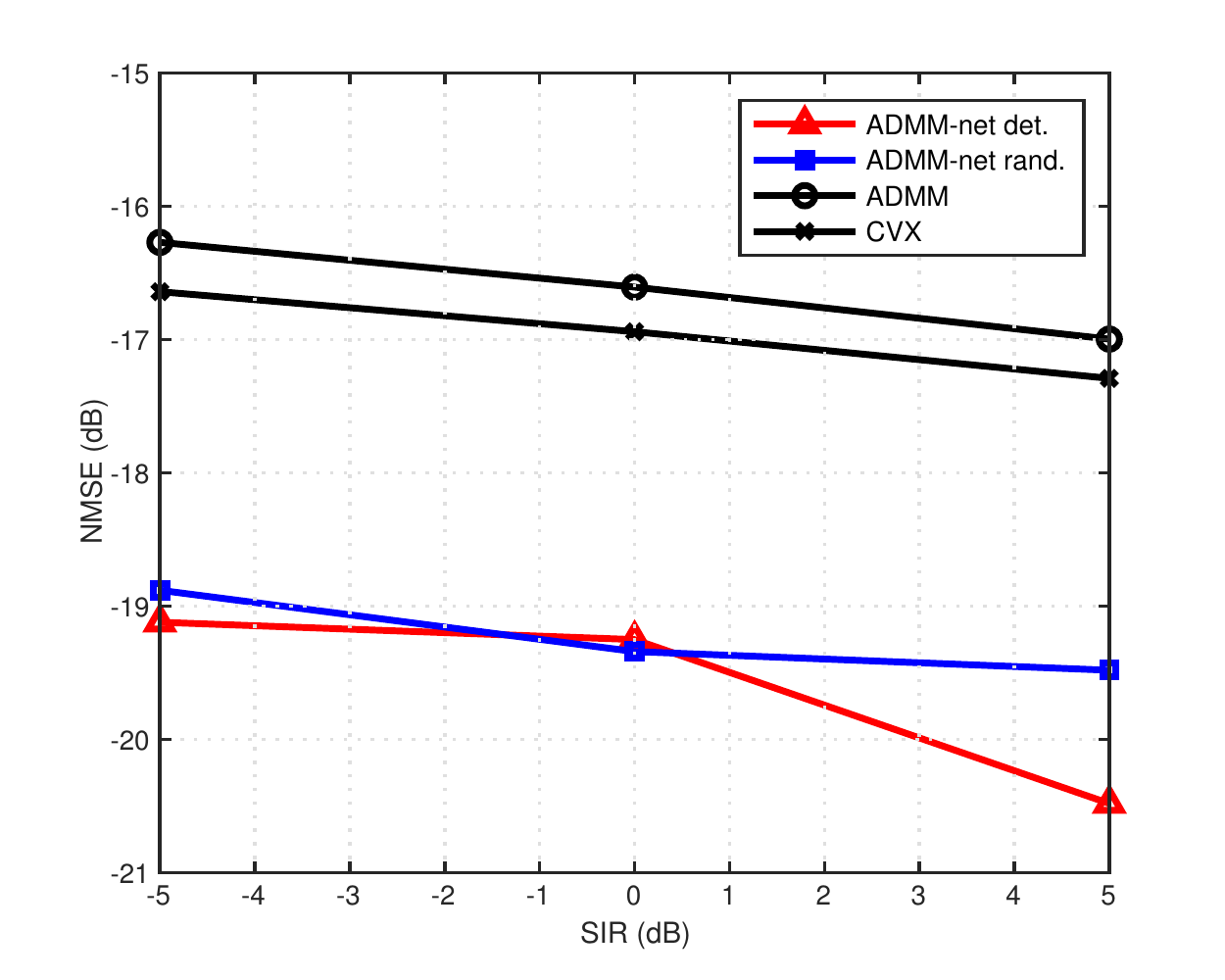}
\caption{ADMM-Net $\mathrm{NMSE}$ versus $\mathrm{SIR}$. 2 scatterers, 25\% spectrum overlap, $\mathrm{SNR}=15$.}
\label{fig:SIR}
\end{figure}

Four networks were trained: three were trained with deterministic $\mathrm{SIR}$s in $\{-5,\,0,\,5\}$, and one was trained on data with random $\mathrm{SIR}$s drawn from $\mathrm{uniform}(-5,\,5)$. For evaluation, we used three test sets with respective $\mathrm{SIR}$s $-5$, $0$, and $5$. Results are plotted in Fig. \ref{fig:SIR}. The red and blue curves are analagous to those in Fig. \ref{fig:SNR-exp}.

\begin{figure}[h]
\includegraphics[width=75mm]{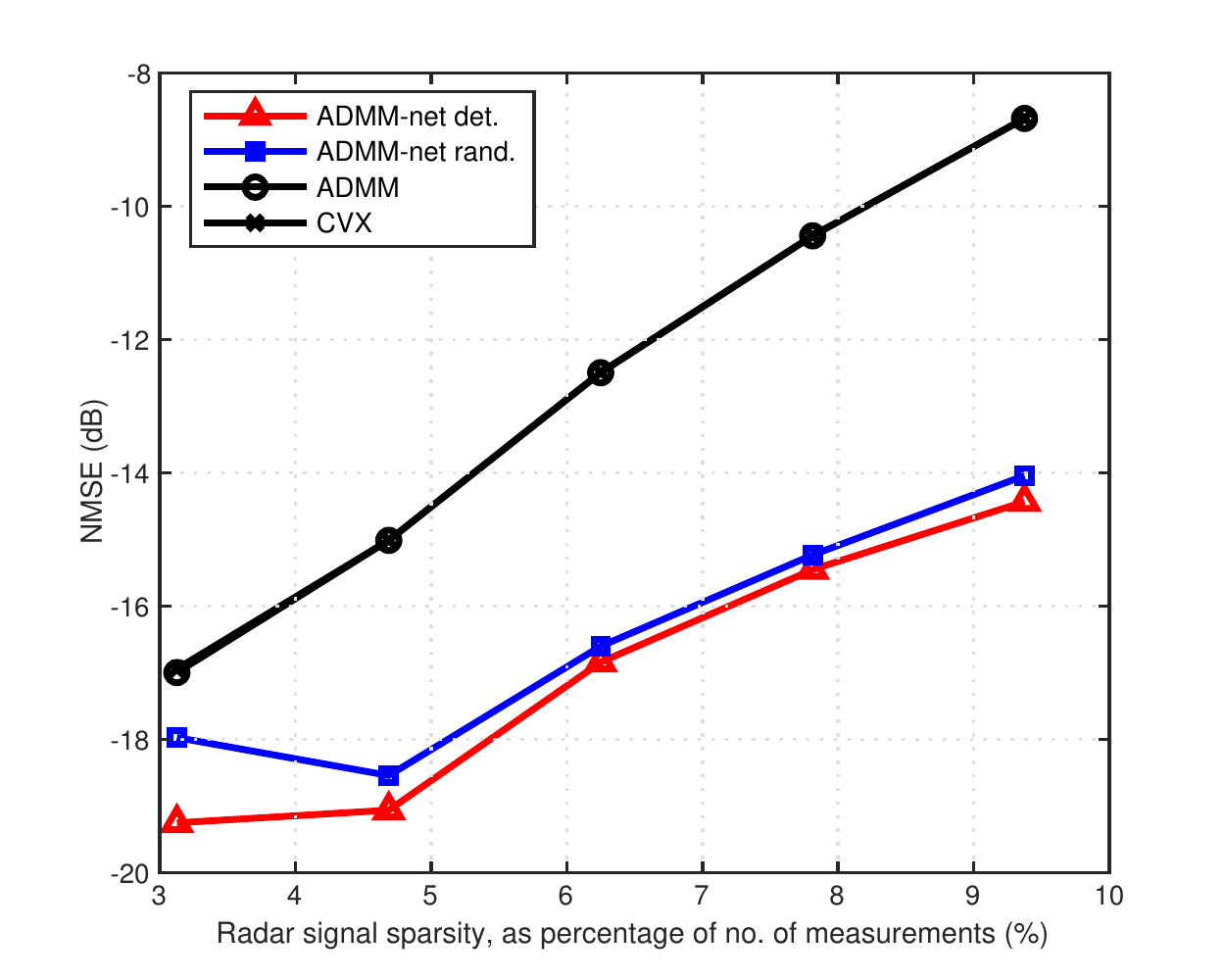}
\caption{$\mathrm{NMSE}$ versus sparsity level where the number of scatterers varies from 2 to 6. 25\% spectrum overlap, $\mathrm{SNR}=15$, $\mathrm{SIR}=0$.}
\label{fig:sparsity-exp-targets}
\end{figure}

\begin{figure}[h]
\includegraphics[width=75mm]{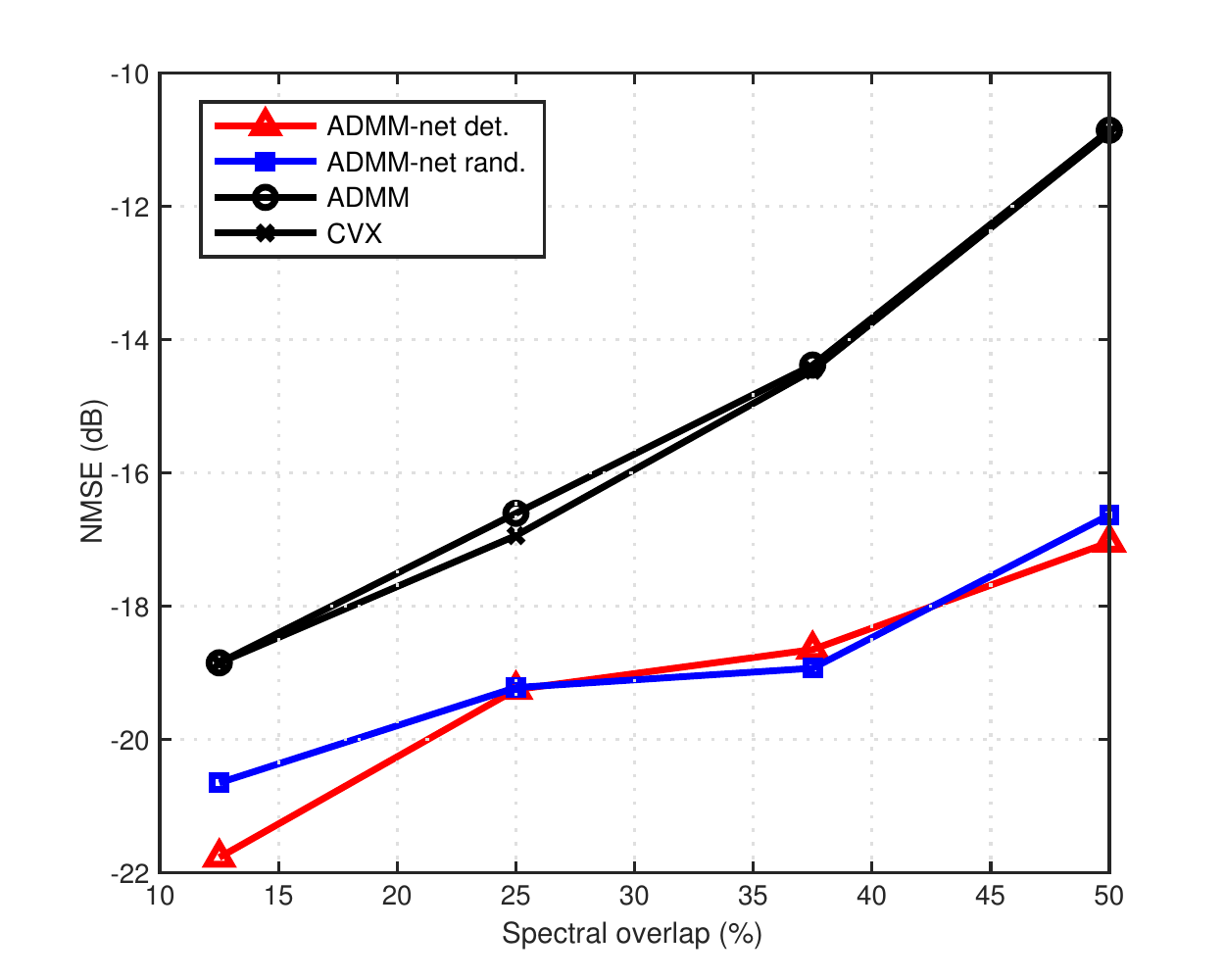}
\caption{$\mathrm{NMSE}$ versus sparsity level where $\|\mathbf b\|_0$ varies from 8 (12.5\% overlap) to 32 (50\% overlap). 2 scatterers, $\mathrm{SNR}=15$, $\mathrm{SIR}=0$.}
\label{fig:sparsity-exp-comm}
\end{figure}

\begin{table}[!htb]
\caption{Run times in milliseconds for the SNR experiments, averaged over 1000 samples.}
\label{tab:runtime}
\centering
\begin{tabular}{c|cccc}  
\toprule
 Method & 5 dB & 10 dB & 15 dB \\
\midrule
ADMM-Net (5 stages) &0.40&0.40&0.40\\
ADMM &22&26&29\\
CVX &510&550&600\\
\bottomrule
\end{tabular}
\end{table}

\begin{figure*}
\centering
	\subfloat[][]{\includegraphics[width=62mm]{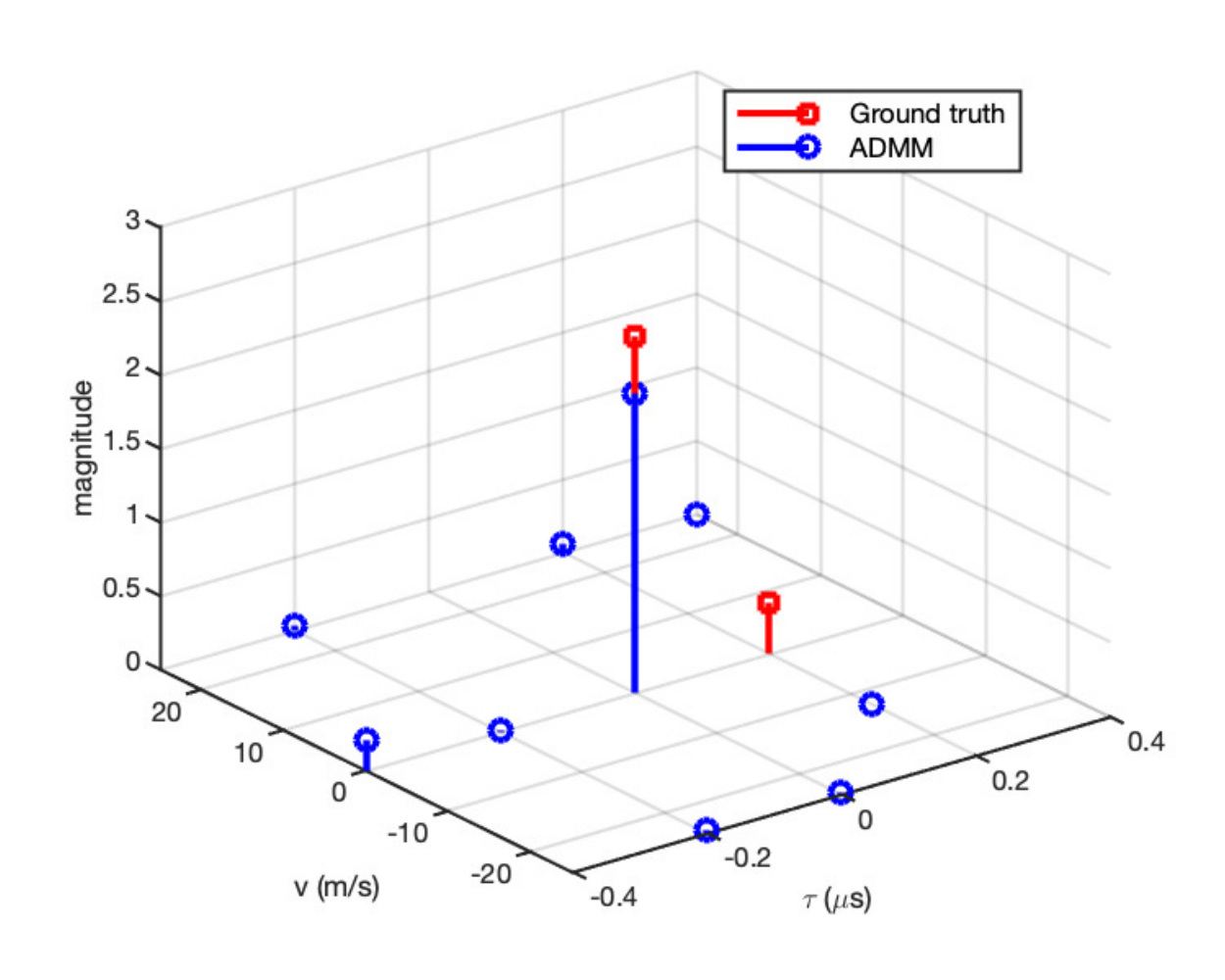}}
	\subfloat[][]{\includegraphics[width=62mm]{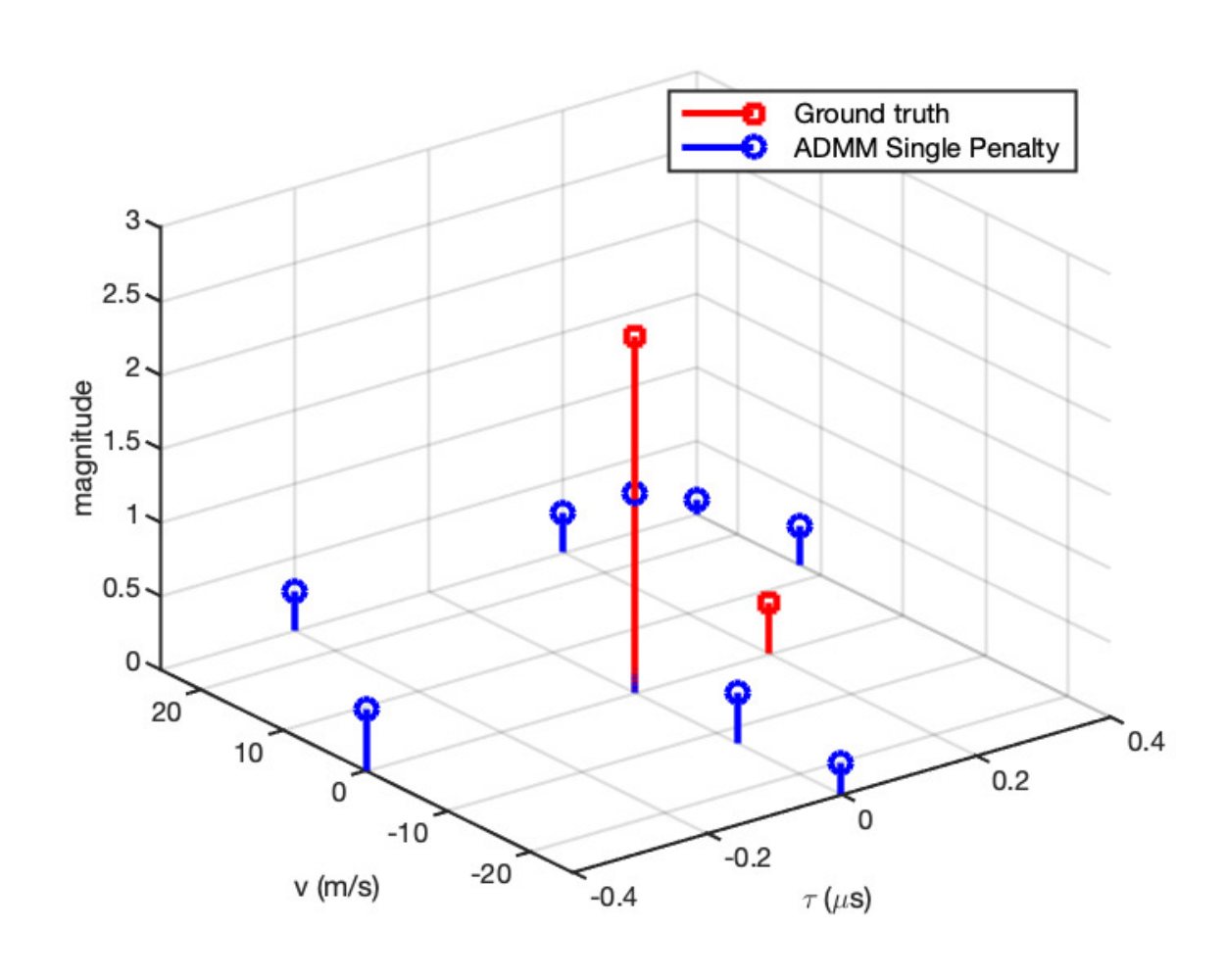}}
	\subfloat[][]{\includegraphics[width=62mm]{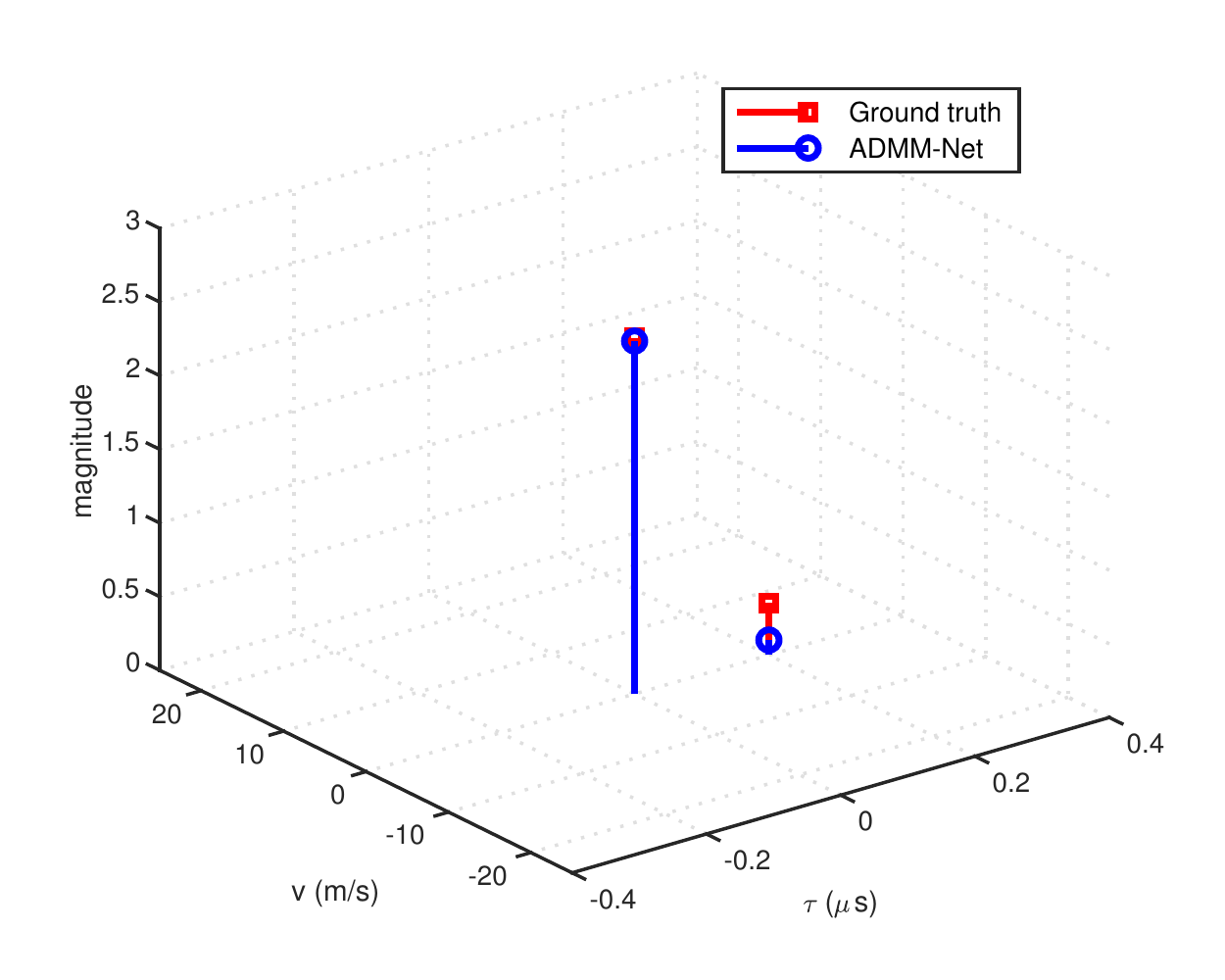}}
	\caption{Recovered range-velocity image slice for (a) ADMM, (b) ADMM w/ single penalty, and (c) ADMM-Net. The two scatterers have magnitudes 2.4 and 0.3 and the same angular position.}
	\label{fig:stemimage}
\end{figure*}

\subsubsection{Sparsity}
For radar sparsity, a total of six networks were trained. Five networks were trained on data sets with deterministic sparsity levels in $\{2,\, 3,\, 4,\, 5,\, 6\}$; within each of the five sets $\|\mathbf w\|_0$ was the same for all samples. One network was trained on data with random sparsity levels, where the sparsity of each sample was drawn from $\mathrm{uniform}(2,\, 6)$. All six sets had $\|\mathbf b\|_0=16$, $\mathrm{SNR} = 15 \ \mathrm{dB}$, and $\mathrm{SIR} = 0 \ \mathrm{dB}$. Note that as the number of scatterers increases, the coefficients must decrease in magnitude in order to yield a given $\mathrm{SNR}$. For evaluation, we fixed $\|\mathbf b\|_0=16$ and varied $\|\mathbf w\|_0$ from $2$ to $6$. Results are plotted in Fig. \ref{fig:sparsity-exp-targets}. Each point on the red curve corresponds to the test set $\mathrm{NMSE}$ of the particular network trained on the (deterministic) sparsity level equal to the point's abscissa. The blue curve plots the $\mathrm{NMSE}$ of the network trained on the data with uniformly distributed sparsity levels. 

Similarly, for interference sparsity, three networks were trained on data sets containing samples with a single deterministic sparsity level belonging to $\{12.5\%,\, 25\%,\, 37.5\%,\, 50\%\}$. The random sparsity level data was generated such that $\|\mathbf b\|_0 \sim \mathrm{uniform}(8,\,32)$. The spectral location and number of interferers were assumed to be the same for each MIMO channel and were allowed to vary from sweep to sweep, but not within a sweep. All four sets had $\|\mathbf w\|_0=2$, $\mathrm{SNR} = 15 \ \mathrm{dB}$, and $\mathrm{SIR} = 0 \ \mathrm{dB}$. Note that as the number of interferers increases, their magnitudes must decrease in order to yield the same $\mathrm{SIR}$. For evaluation, we fixed $\|\mathbf w\|_0=2$ and varied $\|\mathbf b\|_0$ from $8$ to $32$. Results are plotted in Fig. \ref{fig:sparsity-exp-comm}. The red and blue curves are analagous to those in Fig. \ref{fig:sparsity-exp-targets}.

\subsubsection{Recovered Image}
\label{sec:recoveredimage}
To provide a qualitative account of the methods' outputs as well as demonstrate super-resolution capability, we simulate two scatterers in neighboring range grid points and the same velocity-angle grid point. Fig. \ref{fig:stemimage} shows a range-velocity image slice---the slice which corresponds to the scatterers' angle location---for three methods: ADMM, ADMM single-penalty, and ADMM-Net. The respective $\mathrm{NMSE}$s of the (total) recovered images are $-12.0\ \mathrm{dB}$, $-4.7\ \mathrm{dB}$, and $-18.4\ \mathrm{dB}$. In all scenarios, single-penalty ADMM yielded an $\mathrm{NMSE}$ of $-5\ \mathrm{dB}$ or higher, except the scenario $\mathrm{SIR} = 5 \ \mathrm{dB}$ in which the error was $-9\ \mathrm{dB}$.

\subsubsection{Training time}
The 5-stage network training time (45 epochs, $N_\mathrm{train} = 4.5\times10^6$) was approximately 120 minutes on a 2-core server with a single Nvidia Tesla K80. On the same server, the 9-stage network in Fig. \ref{fig:siso_num_stages_experiment} (45 epochs, $N_\mathrm{train} = 5\times10^6$) took approximately 250 minutes to train.

\subsubsection{Run time}
Table \ref{tab:runtime} lists the run times in the SNR experiment, averaged over the test set, for each algorithm, run in Matlab on a MacBook Pro with 8 GB of RAM and a 2.4 GHz Intel i5 processor. The ADMM run time is defined as the time until the convergence criterion
\begin{align}
\label{eq:convergencecriterion}
\frac{\mathrm{NMSE}(k+1) - \mathrm{NMSE}(k)} {\mathrm{NMSE}(k)} < 10^{-6}
\end{align} 
is satisfied, where $\mathrm{NMSE}(k)$ is the $\mathrm{NMSE}$ at iteration $k$. The 5-stage ADMM-Net has a constant run time, equal to the run time of 5 ADMM iterations.

\subsection{Discussion}
\label{sec:discussion}
The deterministically trained ADMM-Nets, tested on data akin to their training sets, outperform ADMM and CVX by at least 2 dB in every scenario, and the performance gap widens to around 4 dB as $\mathrm{SNR}$ decreases below 15 dB, a region of significant practical interest. Moreover, the 5-stage ADMM-Net is between 50 and 80 times faster than ADMM, and between 1275 and 1500 times faster than CVX; see Table \ref{tab:runtime}. Qualitatively, among the recovered images in Fig. \ref{fig:stemimage} ADMM-Net's is the cleanest and most accurate. Also evident from Fig. \ref{fig:stemimage} is the benefit of the two-penalty term optimization objective over the single-penalty objective.

With regard to robustness, we find that the deterministically trained networks are most accurate on test data with the same properties as their respective training sets, as opposed to data with properties different from the training set. The random data-trained networks perform around 1 dB worse than the deterministic data-trained networks, but they are more robust to test set variation. Lower performance may be caused by the fact that, since the training set size is the same as the others, fewer examples from each scenario are represented. Nonetheless, the performance gap shrinks in more challenging environments, i.e. lower $\mathrm{SNR}$, more spectrum overlap, etc.

\section{Conclusion}
We have shown that deep learning, in particular the deep unfolding framework, can significantly improve upon ADMM and CVX for communication interference removal in stepped-frequency radar imaging. The added cost is network training, which can be done in a matter of hours. Training data comes ``for free" by virtue of the statistical signal model, and thus deep unfolding makes fuller use of prior knowledge than standard iterative algorithms, adapting theoretically sound, generally applicable procedures to problem-specific data.

How can we account for the performance ADMM-Net? Certain unfolded networks are designed to learn only algorithm hyperparameters and thus have a clear-cut ``parameter-tuning" interpretation; others, such as our ADMM-Net, learn algorithm \emph{operations}, and thus may elude such a straightforward account. In some cases the learned operations do coincide with those suggested by theory; a VAMP-inspired network, randomly initialized, learns a denoiser matched to the true signal priors \cite{7934066}. ADMM-Net, on the other hand, is initialized as theoretically prescribed, whence it then deviates through training. Further insight might be found in identifying redundancies among the learnable parameters. For example, in LISTA one learnable matrix converged to a final state determined by another, thus allowing a reduction in the number of parameters without altering performance \cite{istanet2018chen}.

\bibliographystyle{IEEEtran}
\bibliography{./database.bib} 
\end{document}